\documentclass[final,3p,times,onecolumn,11pt]{elsarticle}

\usepackage{graphicx}
\biboptions{sort&compress}
\usepackage{amssymb}

\usepackage{lineno}
\usepackage{subeqnarray}
\graphicspath{{./figures/}}

\usepackage{amsmath, subfigure, hyperref, xcolor}

\begin{document}

\begin{frontmatter}


\title{Influence of low frequency modes on dynamical concertedness in double proton transfer dynamics}





\author{Priyanka Pandey}%
\ead{priyanka@iitk.ac.in}
\address{
Department of Chemistry, Indian Institute of Technology \\
Kanpur, Uttar Pradesh 208016, India }%
\author{Shibabrat Naik}%
\ead{s.naik@bristol.ac.uk}
\address{
School of Mathematics, University of Bristol \\
Fry Building, Woodland Road, Bristol BS8 1UG, United Kingdom }%
\author{Srihari Keshavamurthy\corref{cor1}}%
\ead{srihari@iitk.ac.in}
\address{
Department of Chemistry, Indian Institute of Technology \\
Kanpur, Uttar Pradesh 208016, India }%

\begin{abstract}
We analyze the classical phase space dynamics of a three degree of freedom Hamiltonian that models multiple bond breaking and forming reactions. The model Hamiltonian, inspired from studies on double proton transfer reactions, allows for exploring the dynamical consequences of higher index saddles on multidimensional potential energy surfaces. Studies have shown that coupling of low frequency transverse modes to the reaction coordinate can significantly influence the reaction mechanism, concerted or sequential, as inferred from a reduced dimensional analysis. Using the notion of dynamically concerted and sequential pathways,  we provide insights into the role of the transverse modes by studying the delay times between the formation of two bonds. The delay time distribution, used extensively in earlier studies, is placed on a firm dynamical footing by correlating it with the phase space manifolds, determined using the technique of Lagrangian descriptors.  We establish the utility of Lagrangian descriptors in identifying the phase space manifolds responsible for the dynamically concerted and dynamically sequential pathways. 
\end{abstract}


\end{frontmatter}

\linenumbers

\section{Introduction}
\label{sect:intro}

The theory of nonlinear dynamical systems is a natural framework for understanding chemical reactions. There are several reasons for such a claim, but two of them are key. Firstly, breaking of a bond is only possible if the vibrations are modeled as nonlinear oscillators. Secondly, the canonical paradigm of associating an energized  molecule with many such nonlinear oscillators that are coupled together~\cite{uzer1991theories,logan1990quantum,gruebele2003mechanism,leitner2015quantum} leads to a  rich and complex dynamical behaviour that necessitates a phase space perspective for proper analysis and interpretation~\cite{keshavamurthy2013scaling,karmakar2020intramolecular}.  Indeed the central notion of a transition state is best understood as a dynamical bottleneck that is formed by certain invariant manifolds in the phase space~~\cite{waalkens2010geometrical,wiggins2016role,pollak1978transition,pechukas1979classical,waalkens2007wigner,uzer2002geometry,jaffe2005new}. Reaction rates can then be associated with fluxes through appropriate bottlenecks~\cite{pechukas1976statistical,miller1998spiers,pollak2005reaction}.  Thus, a local dynamical perspective on Transition state theory (TST) has provided fresh insights into the usefulness and limitations of TST in the microcanonical~\cite{ezra2009microcanonical}, canonical~\cite{collins2010phase} and more general~\cite{bartsch2005transition,kawai2009dynamic1,kawai2009dynamic2,feldmaier2020influence,cciftcci2013reaction} settings.    

Apart from the rates, there is yet another important aspect of a chemical reaction that is enshrined in the TS - the mechanism. In fact, identifying the correct TS is essentially equivalent to a knowledge of the mechanism of the reaction.  For reactions involving a single TS (elementary reactions)  one therefore associates a single mechanism that leads to the transformation of the reactants to products.  However, several reactions are associated with potential energy surface (PES) that exhibit novel features~\cite{tantillo2018applied} like extended flat regions (calderas or more generally entropic intermediates), ambimodal TS, valley ridge inflection points, several distinct saddle points (multiple TSs), and saddle points with more than one unstable direction (higher index saddles). It is now clear that the existence of such features on the PES can lead to significant dynamical effects. Examples include dynamical matching~\cite{carpenter1985trajectories,carpenter1995dynamic,collins2014nonstatistical,katsanikas2018phase,katsanikas2020dynamical,geng2021influence}, nonstatistical branching ratios~\cite{rehbein2011we,collins2013nonstatistical,thomas2008control}, energy dependent product selectivity~\cite{kurouchi2018labelling,kurouchi2016controlling,quijano2011competition}, and switching of reaction mechanisms~\cite{pandey2021classical,teramoto2011dynamical,accardi2010synchronous}.  Consequently, there is an increased focus now on trajectory-based analysis of complex reactions. 

In the current work we are interested in understanding the dynamics of
reactions that involve breaking and forming of multiple bonds. Here one invariably has to face up to a fundamental and essential mechanistic question: is the process occurring sequentially or in a concerted fashion? In this regard, the Diels-Alder and the double proton transfer (DPT) reactions have provided a rich arena  to explore the role of dynamics in determining the correct reaction mechanism~~\cite{goldstein1996density,pham2014diels,houk2020evolution,black2012dynamics,takeuchi2007answer,accardi2010synchronous,ushiyama2001successive,homayoon2014calculations,abdel2011laser,abdel2010infrared}.  From a fundamental point of view the possibility of more than one distinct pathway is linked with the presence of  several distinct TSs. Although traditionally one associates TS with a index-$1$ saddle point on the multidimensional PES, several studies indicate that the dynamical influence of higher index saddles on the PES can also be a deciding factor in identifying the dominant mechansim~\cite{nagahata2013reactivity,pradhan2019can,lu2014evidence,quapp2015embedding,rashmi2021second}.  For instance, for energies above the index-$2$ saddle one can have a time-dependent switching between the concerted and sequential pathways. Recently, it was shown~\cite{pandey2021classical} that such a dynamical mechanism switch is an inherently classical phenomenon. Moreover, owing to the mixed regular-chaotic nature of the classical phase space, initial quantum wavepackets that are centered at specific regions of the classical phase space can undergo strikingly different mechanism-switching dynamics~\cite{pandey2021classical}. Interestingly, the switching timescale is typically of the order of a  bond stretching time period and hence ultrafast. These observations therefore raise questions on the utility of a purely non-dynamical classification of the mechanism as concerted or sequential. Such concerns have been raised by Carpenter in his early work on the dynamic matching phenomenon wherein he emphasizes the ``hazards associated with partitioning of mechanisms into stepwise and concerted categories" based purely on the features on the static potential energy surface~\cite{carpenter1995dynamic}. More recently,  Houk and coworkers introduced quantitative measures for classifying the mechanism as dynamically concerted or sequential. Thus, for a given trajectory, if the time delay between the formation of the first bond and the second is shorter than a specified timescale then that particular trajectory is classified as dynamically concerted~\cite{black2012dynamics}. As a consequence the central quantity of interest is the distribution of the delay times associated with an appropriate ensemble of trajectories. Depending on the nature of the delay time distributions one can identify the mechanism as dynamically concerted or sequential. 

Note that the approach of Houk and coworkers~\cite{black2012dynamics} implicitly invokes the dynamics in the full classical phase space. Understandably, a detailed phase space analysis of the ab initio molecular dynamics based studies of reactions like the Diels-Alder is far from easy. At the same time, there is no denying the fact that rationalizing the dynamics based on the phase space structures is expected to yield rich dividends in terms of our ability to predict rather than simply observe or compute. Thus, from a nonlinear dynamical systems point of view it is natural to expect that the delay time distributions are intimately linked to the disposition of the stable and unstable phase space manifolds  that lead to transport from the reactant to the product regions. However, identifying, let alone computing, such manifolds in very high dimensions is not feasible at the present moment. A crucial question then is this: can reduced dimensional models capture enough of the essential dynamics to allow for at least  qualitative predictions? 
The answer, as apparent from the several studies utilizing ``minimal" models, is yes. For example, significant dynamical insights into the phenomenon of roaming and dynamic matching have come from phase space studies on the low dimensional model systems~\cite{mauguiere2014multiple,mauguiere2015phase,mauguiere2016phase,mauguiere2017roaming,montoya2020revealing,carpenter1985trajectories,carpenter1995dynamic,collins2014nonstatistical,katsanikas2018phase,katsanikas2020dynamical}. Nevertheless, the detailed study of a electrocyclic ring opening reaction by Kramer et al. highlights the central issues in this regard~\cite{kramer2015reaction}. A comparison of the direct dynamics calculations (in a $36$-dimensional phase space) with the reduced two-dimensional model dynamics for the same reaction revealed that the two do share dynamical similarities. However, they make an important point - in the event that large amplitude modes, which would be considered as "spectator" modes in the reduced dimensional treatment, couple to the reaction coordinate, the  dynamics may be more complicated then what would be predicted by the reduced dimensional models. Note that one can associate large amplitude modes with low frequency vibrations and in a molecule with symmetry the various low frequency modes can couple to the reactive mode in different ways due to the symmetry constraints. Thus, apart from  leading to a more complicated dynamics, can the coupling of specific low frequency modes alter the inferred reduced dimensional mechanism itself?  

Since the present study focuses on the DPT reaction in a specific class of molecules, we mention a few examples from  earlier studies that highlight the importance of the low frequency modes.  In their extensive review of multiple proton transfer dynamics, Smedarchina et al.~\cite{smedarchina2006multiple} have argued for the importance of the coupling of low frequency skeletal vibrations to the proton transfer modes. Furthermore, in a path integral molecular dynamics simulation Yoshikawa et al. have shown that the low frequency out-of-plane vibration can suppress the concerted pathway in porphycene molecule~\cite{yoshikawa2010theoretical,yoshikawa2012quantum}. Another example comes from the Car-Parrinello molecular dynamics study of porphycene by Walewski et al., where it was observed that excitations of selective low frequency modes, and combinations thereof, tend to enhance or suppress the different mechanisms~\cite{walewski2010car}. It is also relevant to point out the classical ab initio molecular dynamics study of DPT by Ushiyama and Takatsuka where, apart from hints to the importance of delay time distributions, the crucial role of skeletal vibrations to the second proton transfer was emphasized~\cite{ushiyama2001successive}. 

Clearly, and as discussed in detail in sec.~\ref{sec:modham}, for a DPT reaction with two reactive modes, coupling of even one low frequency transverse mode results in a system with three degrees of freedom. In this work we investigate the classical dynamics  of such a model system with the aim of explicitly correlating the delay time distributions with the appropriate phase space manifolds. In particular, as mentioned above, we investigate the  influence on the delay time distributions due to the coupling of a third mode belonging to a specific symmetry class. The results show that while high frequency modes do not significantly change the fraction of concerted trajectories, the low frequency modes can substantially reduce the fraction.  An explanation of our delay time results in terms of the phase space structures is given by computing the appropriate manifolds using the technique of Lagrangian descriptors(LD). We show that the the LD maps faithfully capture the changes in the delay time distributions with varying coupling strength of the third mode.  In sec.~\ref{sec:modham} and ~\ref{1DHamiltonian} we motivate the model Hamiltonian used in our study. The influence of the third mode on the delay time distributions are presented in sec.~\ref{subsect:delaytime}, followed by the results of the delay time distributions for varying frequencies and coupling strengths. In sec.~\ref{subsect:LD_delaytime}, aided by the stability analysis of the linearized flow in ~\ref{jacobian_vf}, the dynamical trajectory observations are correlated with the LD-based determination of the relevant phase space manifolds. Finally, sec.~\ref{sect:conc} concludes with a brief summary and future outlook.  

\section{Model Hamiltonian}
\label{sec:modham}


\begin{figure}[htbp]
\centering
\includegraphics[width=0.75\linewidth]{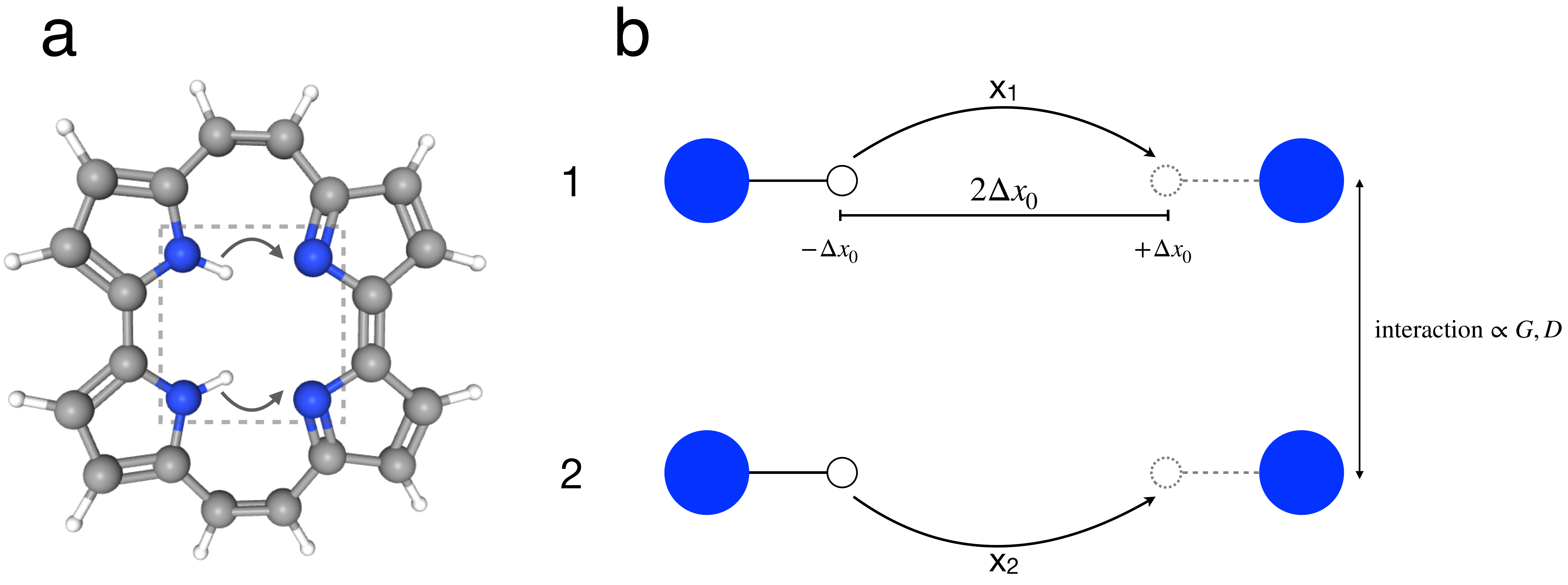}
\caption{(a) The porphycene molecule with $N=38$ atoms. The nitrogen, carbon, and hydrogen atoms are indicated in blue, grey, and white colors respectively. (b) Schematic for the double proton transfer model (indicated by arrows in both the panels) corresponding to the dashed square region shown in (a). The coordinates $(x_1,x_2)$ correspond to the two proton transfer events. Each proton transfer subsystem is described by an appropriate double well potential with minima at $\pm \Delta x_{0}$. Interaction between the two subsystems is mediated by the coupling constant $G$ and $D$ at the leading order. See the main text and \ref{1DHamiltonian} for details. }
\label{Fig1}
\end{figure}

\begin{figure}
 \centering
 \includegraphics[width=01.0\textwidth]{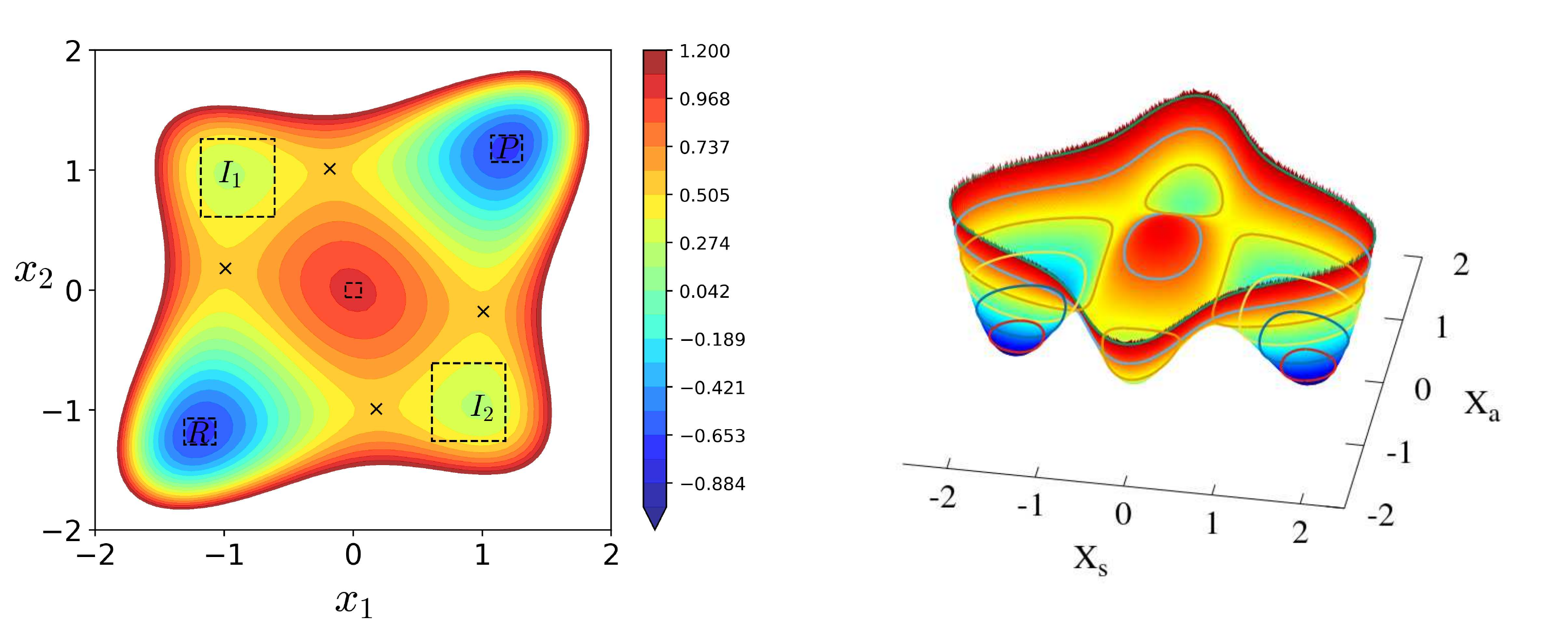}
 \caption{(Left panel) A contour plot of the model potential energy surface in the local ($x_1,x_2$) coordinates. (Right panel) A three dimensional plot of the potential energy surface in the mass-weighted ($X_s,X_a$) coordinates. The parameter values are $G = 0.2$, and $D=0.15$. Different regions of the potential energy surface are shown inside the dotted space of the contour plot where $R$, $P$, and $I_{1,2}$ represent the reactant, product and intermediate regions respectively. The central box represents the index-$2$ saddle region. Note that the PES also exhibits four index-$1$ saddles, which are shown as $\times$.
}
 \label{fig:2Dpes}
\end{figure}

To study the double proton transfer reaction, we consider a model three degree of freedom Hamiltonian motivated  by the models introduced by Smedarchina et al.~\cite{smedarchina2007correlated,smedarchina2018entanglement} in their extensive studies.
In Fig.~\ref{Fig1} we show a schematic for the DPT process. The two proton transfer events occurring in the molecule  (labeled as subsystem $1$ and $2$ in the figure) are described by one dimensional coordinates $x_1$ and $x_2$ with associated masses $m_1$ and $m_2$, which are taken to be equal to the proton mass $m_{H}$. As shown in detail in \ref{1DHamiltonian}, an appropriate model two degrees of freedom dimensionless Hamiltonian for the coupled proton transfer is conveniently expressed  in terms of the coordinates $(X_s,X_a) \equiv (\sqrt{M} x_s,\sqrt{M} x_a)= (\sqrt{M}(x_1+x_2)/2,\sqrt{M}(x_1-x_2)/2)$ with $M = m_1 + m_2 = 2 m_{H}$. The Hamiltonian is of the form  
\begin{equation}
    H({\bf X},{\bf P}) = \frac{1}{2} (P_{s}^{2} + P_{a}^{2})  + U(X_s,X_a)
    \label{massscaledham2D}
\end{equation}
with $(P_s,P_a)$ being the momenta conjugate to $(X_s,X_a)$. The two dimensional potential energy surface is given by
\begin{equation}
    U({\bf X}) = \bar{\alpha}_{s} \left[X_{s}^{2} - (\Delta X_s)^{2}\right]^{2} + \bar{\alpha}_{a} \left[X_{a}^{2} - (\Delta X_a)^{2}\right]^{2} + 2\bar{R} X_{s}^{2} X_{a}^{2} + {\cal U}(G,D)
    \label{massscaledpot2D}
\end{equation}
In the above we have denoted $({\bf X},{\bf P}) \equiv (X_s,X_a,P_s,P_a)$ with the parameters $\bar{\alpha}_{s} = \bar{\alpha}_{a} \equiv (1-D)/M^{2}$ and $\bar{R} \equiv (3+D)/M^{2}$.   The various minima on the PES are given in terms of the quantities
\begin{equation}
 \Delta X_{s,a} = \sqrt{\frac{M(1\pm G)}{1-D}} 
\end{equation}
and the constant energy shift is denoted as
\begin{equation}
    {\cal U}(G,D) = 1 - \bar{\alpha}_s (\Delta X_s)^{4} - \bar{\alpha}_a (\Delta X_a)^{4}
\end{equation}
The parameters $G$ and $D$ are specific to a given system (molecule) and correspond to the coupling of the two proton transfer coordinates. We refer the reader to the ~\ref{1DHamiltonian} for a detailed derivation of the above Hamiltonian along with the relevant mass, length, and time scales. In Fig.~\ref{fig:2Dpes} the two dimensional PES are shown in the two different sets of coordinates. Note that, in general, the number and type of critical points on the PES~\cite{smedarchina2018entanglement} depend on the values of $G$ and $D$, mimicking a wide variety of dynamical systems. For the values of interest to us in the current work the PES exhibits a total of nine critical points which, as seen in Fig.~\ref{fig:2Dpes}, include four minima, four index-$1$ saddles and one index-$2$ saddle.

In the molecular context the Hamiltonian in Eqn.~\ref{massscaledham2D} captures the dynamics corresponding to the key reactive degrees of freedom. Thus, for an $N$-atom molecule of interest with $(3N-6)$ vibrational degrees of freedom Eqn.~\ref{massscaledham2D} accounts for two of the degrees of freedom. However, the remaining $(3N-8)$ modes that are transverse to the reactive modes typically do couple to the ${\bf X}$ degrees of freedom to varying extents. Moreover, if the molecule of interest has a certain point group symmetry then the various transverse modes are constrained to couple to ${\bf X}$ with specific functional form of the coupling potentials. For instance, in the context of DPT all the $(3N-8)$ modes denoted by ${\bf Y}$ couple at leading order via the potential
\begin{equation}
    U_{\rm coup}({\bf X},Y_k) = \frac{1}{2}\omega_{kY}^{2}\left[Y_k - \frac{\lambda}{\omega_{kY}^{2}} g({\bf X})\right]^{2}
    \label{eq:pot3Dcoup}
\end{equation}
with $g({\bf X})=X_{s,a}, X_{s} X_{a}$, and $X_{s,a}^{2}$ depending on the symmetry class to which the $Y_k$-mode belongs. In Eqn.~\ref{eq:pot3Dcoup} the $Y_k$ mode is modeled as a harmonic oscillator with the frequency of the mode  denoted by $\omega_{kY}$ and $\lambda$ being a measure of the coupling strength. It is interesting to note that with the above form of coupling one can still think of the ${\bf Y}$-modes as providing a ``bath", albeit a structured one. This is in contrast to the usual system-bath models wherein all the ``bath" ${\bf Y}$-modes would couple bilinearly with a specified spectral density. 
Clearly, the dynamical implications of a structured and a non-structured bath are expected to be quite different for the reaction process.

As mentioned in the introduction, several studies have indicated the importance of including the additional modes since they can have a significant effect on the mechanism inferred from an analysis of  the reduced dimensional Hamiltonian. Thus, certain symmetry modes tend to enhance a specific mechanism (concerted or sequential) whereas certain other symmetry modes act in an opposite manner~\cite{walewski2010car}. Therefore, in order to rationalize the observed DPT rates one minimally needs to include  two of the ${\bf Y}$-modes, resulting in a four degree of freedom Hamiltonian. However, understanding global phase space transport and linking it to the dynamical influence of the coupled ${\bf Y}$-modes in this case is a challenging task. Instead, here we address a simpler yet nontrivial question - can the coupling of a ${\bf Y}$-mode with a given symmetry significantly influence the reaction mechanism as inferred from the low dimensional system in Eqn.~\ref{massscaledham2D}? And, if so, what is the dynamical origin of such a modulation?   As  noted above  in Eqn.~\ref{eq:pot3Dcoup}, there are several choices for the model Hamiltonian according to the $g({\bf X})$ of interest. In this work we focus on the specific three degree of freedom Hamiltonian 

\begin{eqnarray}
    H({\bf X},{\bf P},Y,P_Y) &=& \frac{1}{2} (P_{s}^{2} + P_{a}^{2} + P_{Y}^{2})  + U({\bf X})
    + \frac{1}{2} \omega_{Y}^{2} \left[Y - \frac{\lambda}{\omega_{Y}^{2}} X_{s}^{2} - \frac{\lambda'}{\omega_{Y}^{2}} X_{a}^{2} \right]^{2} \\
    &\equiv& \frac{1}{2} (P_{s}^{2} + P_{a}^{2} + P_{Y}^{2})  + V({\bf X},Y) 
    \label{massscaledham3D}
\end{eqnarray}
with $U({\bf X})$ being the potential in Eqn.~\ref{massscaledpot2D} and we continue to adopt the mass-weighted coordinate representation. Note that the above form of coupling corresponds to the so called $a_{g}$-symmetry $Y$-mode and the importance of this coupling to DPT has been noted in several earlier studies. For example, in the $N=38$ atom porphycene molecule one has a total of $3N-8=106$ ${\bf Y}$-modes. Among these modes the ones with $a_{g}$ symmetry have substantial projection~\cite{smedarchina2014tunneling} onto the two reactive modes ${\bf X}$.

\begin{figure}
    \centering
    \includegraphics[width=0.6\textwidth]{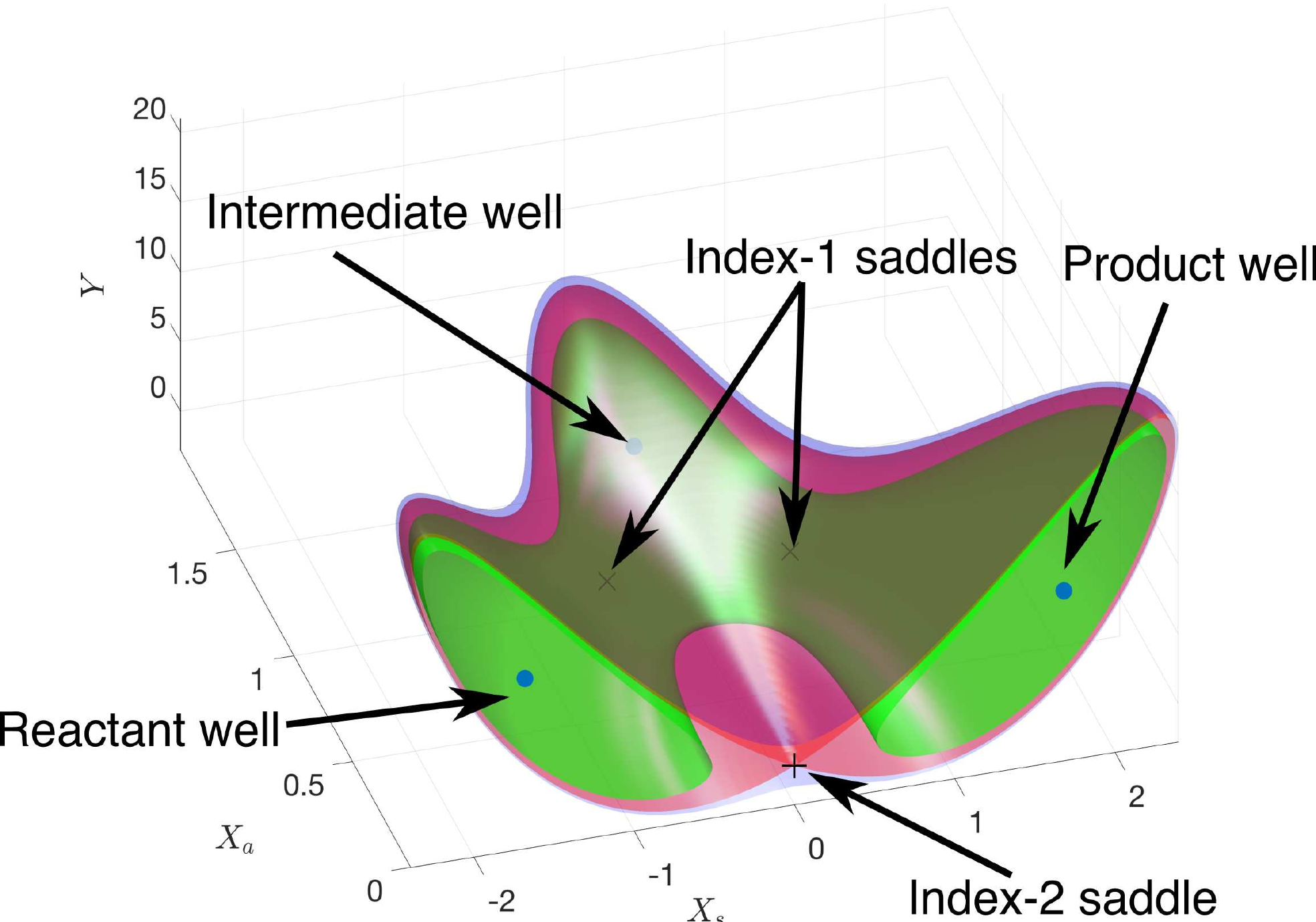}
    \caption{Potential energy visualized as equipotential surfaces for three energies. The green surface denotes $E < E_s$, red surface denotes $E = E_s$, and the blue surface denotes $E > E_s$. The energy of the index-2 saddle is $E_s$ and denoted by the $+$, while the index-1 saddles are shown as $\times$. The parameter values are $G = 0.2$, $D=0.15$, $\omega_{Y} = 0.3$, $\lambda = 0.3$ and $\lambda' = 0.1$ for the total energy, $E = 1.1$ to visualize the equipotential surfaces.}
    \label{fig:pes_3d}
\end{figure}

For the purpose of the current study we choose $0 < G < 1/2$ and $|D| < 2G$ which yields a total of nine critical points.  The details associated with the critical points are given in Table~\ref{tab:eqpt_energy_stab_scaledham3dof}. Note that there is no restriction on the sign of $D$. At the same time, for a given $G$, the dynamics corresponding to positive or negative $D$ can be sufficiently different. For the rest of the paper we fix the values $G = 0.2$ and $D = 0.15$. Moreover, we fix the total energy at $E = 1.1$, which is slightly above the index-$2$ saddle energy (cf. Table~\ref{tab:eqpt_energy_stab_scaledham3dof}). Consequently, both the concerted and sequential pathways (examples can be seen in Fig.~\ref{fig:sampletraj_delyatimes}) from reactant to product are available classically.
A key objective of the current study is to relate the phase space dynamics of the Hamiltonian in Eqn.~\ref{massscaledham3D} with the mechanism of DPT. In particular, we intend to assess the influence of the transverse $Y$-mode with both low and high frequencies $\omega_{Y}$ over a range of the couplings $(\lambda, \lambda')$. In Fig.~\ref{fig:pes_3d} a representation of the PES $V(\mathbf{X},Y)$ is shown for three values of the energy, $E < E_s, E = E_s, E > E_s$, where $E_s$ is the energy of the index-2 saddle at the origin. 

\begin{table}[!ht]
	\centering
	\begin{tabular}{| c | c | c | c |}
		\hline 
		Configuration space coordinates  & Total energy & Linear stability & Description\\
		\hline \hline
				$\left(\pm \Delta X_s, 0, \dfrac{\lambda}{\omega_Y^2}(\Delta X_s)^2 \right)$ & $1 - \bar{\alpha}_s (\Delta X_s)^4$ & C-C-C & {\rm reactant and product} $R,P$\\
		\hline
		$\left(0,\pm \Delta X_a, \dfrac{\lambda^\prime}{\omega_Y^2}(\Delta X_a)^2 \right)$ & $1 - \bar{\alpha}_a (\Delta X_a)^4$ & C-C-C & {\rm intermediates} $I_{1,2}$\\
		\hline
		$\left(X_s^{\ddagger}, X_a^{\ddagger}, Y^{\ddagger} \right)$ & $H^{\ddagger}$ & S-C-C & index-$1$ saddles\\
		\hline
		$(0,0,0)$ & $E_s = 1$ & S-S-C & index-$2$ saddle\\
		\hline
	\end{tabular}
	\vspace{5pt}
	\caption{Equilibrium points, their respective energies and linear phase space stability. The stability types are denoted by $C$ for center and $S$ for saddle. For an explicit expression for $\left(X_s^{\ddagger}, X_a^{\ddagger}, Y^{\ddagger} \right)$ and the associated energy $H^{\ddagger}$ see ~\ref{jacobian_vf}.}
	\label{tab:eqpt_energy_stab_scaledham3dof}
\end{table}

\section{Results and Discussions}
\label{sect:results}

\subsection{Computational preliminaries: defining the initial ensemble, relevant regions, and delay time}
\label{subsect:compdetails}

In order to understand the influence of the $(\lambda,\lambda')$ couplings  and the frequency $\omega_Y$ of the transverse $Y$-mode on the dynamics of the Hamiltonian in Eqn.~\ref{massscaledham3D}, we compute the delay time distribution.  The concept of delay time distribution is motivated by the dynamical studies of Diels–Alder reactions by Houk and coworkers~\cite{black2012dynamics}.
As the name suggests, the delay time corresponds to the time difference between the transfer of the first proton and the subsequent transfer of the second proton. By definition, the delay time is zero for a pure concerted pathway that proceeds directly from the reactant $R$ to the product $P$ via the index-$2$ saddle, and without visiting the intermediate regions. On the other hand, a sequential pathway from $R$ to $P$ via any or both the intermediate regions $I_{1,2}$ yields a finite value for the delay time. Therefore, demarcating different regions in the configuration space (i.e., reactant, intermediate, product, and index-2 saddle) is necessary  in order to  compute the delay time and related measures. Consequently, in Fig.~\ref{fig:2Dpes}, the definitions of the various regions of the PES used in this study are shown as boxes of various sizes. We have divided the PES into five regions. The product and the reactant regions are centered at the minima ($\pm \Delta X_s, 0$), and the region boundaries are at ($\pm 0.1\Delta{X_s}, \pm0.15$) from the minima. Similar regions are defined around the intermediate minima ($0,\pm \Delta{X_a}$). However, for the intermediate regions, the region boundaries are at ($\pm0.40, \pm 0.33\Delta{X_a}$) from the minima. Since the height and the spread of the wells containing the global and local minima vary, therefore we are using different box sizes. The index-$2$ saddle region is centered at (0,0) with the boundaries at ($\pm 0.05,\pm 0.05$).
Note that there are other ways to assign the different regions on the PES and, naturally, the quantitative delay time distributions will be sensitive to the specific choice. However, within reasonable definitions of the regions, at a qualitative level the results are not expected to be significantly different i.e., the key trends with varying parameters are preserved.

\begin{figure}
    \centering
    \includegraphics[width=1.0\textwidth]{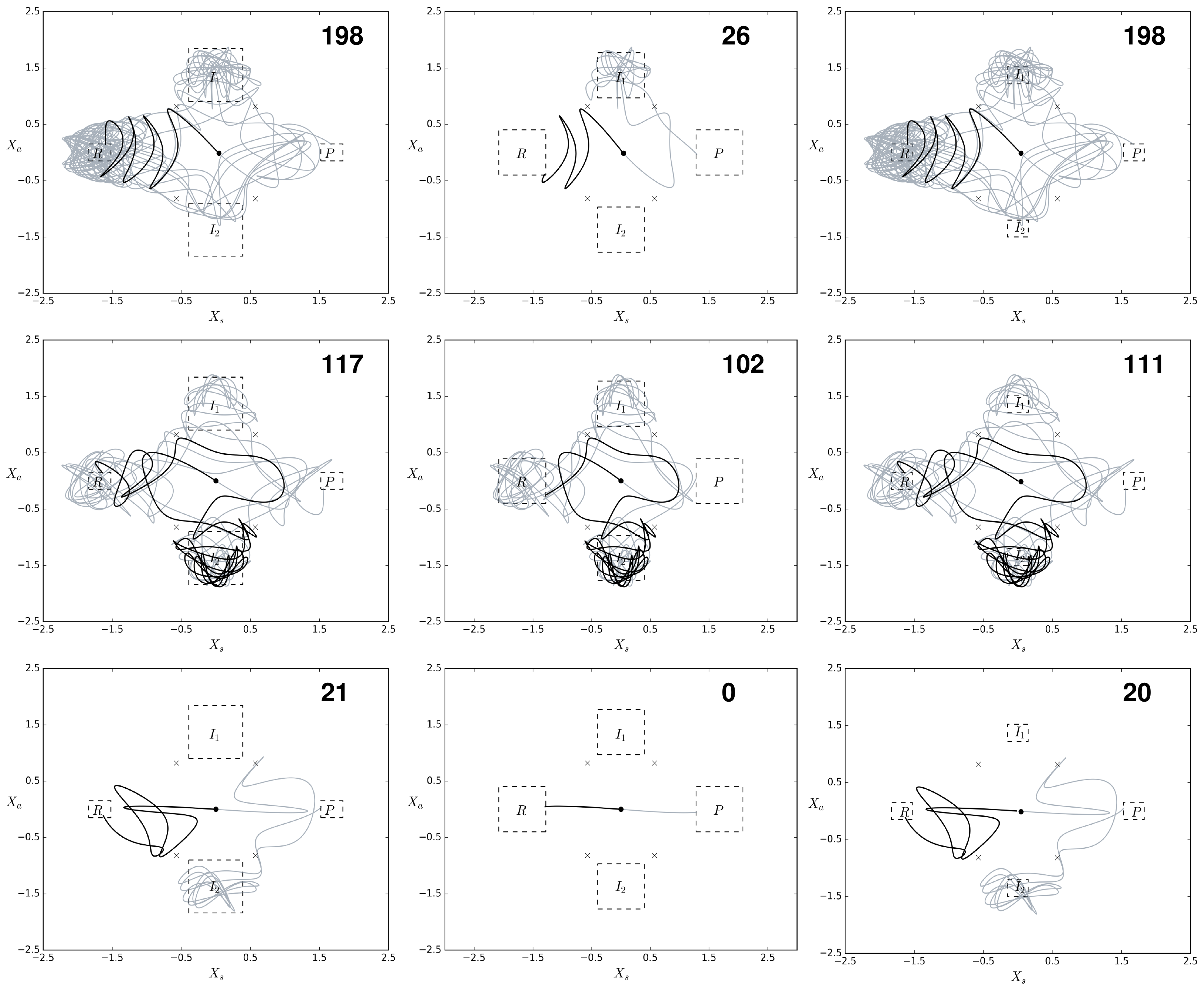}
    \caption{Examples trajectories with different delay times projected onto the $(X_s,X_a)$ space. The various region definitions are shown as dashed boxes. (Left column) Region definitions used in the present work. (Middle column) Increased $R$ and $P$ region sizes. (Right column) Decreased $I_{1}$ and $I_{2}$ region sizes. All trajectories start at the index-$2$ saddle (solid circle) with a total energy $E = 1.1$. The forward and backward time propagation from the index-$2$ saddle are shown in grey and black respectively. The delay time assigned in each case is indicated in the respective panels. Note that the four index-$1$ saddles are indicated by $\times$. The example with zero delay time is a ``pure" concerted trajectory.}
    \label{fig:sampletraj_delyatimes}
\end{figure}

For our calculations, we choose the initial values of position coordinates $X^{(0)}_{s}$ and $X^{(0)}_{a}$ randomly from the index-$2$ saddle region.  This choice of the initial ensemble is to focus on the influence of the index-$2$ saddle on the DPT. In addition, we randomly choose the initial momentum $P^{(0)}_{s} > 0$ and fix the initial value of the third mode coordinate $Y^{(0)}$ and its conjugate momentum $P^{(0)}_{Y}$ at $(0,0)$. Finally, the initial momentum $P^{(0)}_{a}$ is obtained by the energy conservation condition i.e., $H({\bf X}^{(0)},{\bf P}^{(0)},Y^{(0)}=0,P^{(0)}_{Y}=0) = E = 1.1$. The specific total energy value, fixed for the rest of the study, corresponds to being just above the index-$2$ saddle energy.  Furthermore, note that the choice $P^{(0)}_{s} > 0$ corresponds to the trajectories at the index-$2$ saddle having momentum in the direction of the product $P$. It is well known that the product selectivity of a chemical reaction in a trajectory calculation is strongly associated with the momentum distribution at the TS~\cite{wang2009recrossing,carpenter1985trajectories}. Thus, although  different initial momentum distributions at the index-$2$ saddle can lead to quantitatively different results, we believe that the qualitative insights are fairly robust. 
A total of $10^{4}$ trajectories were initiated from the index-$2$ saddle region and propagated both in the forward and backward direction until they reach the product and the reactant regions respectively. Trajectories, propagated up to a final time $t_f = 300$, are deemed to be reactive if they form the product in the forward direction and the reactant in the backward direction.   Depending on the path a reactive trajectory takes and the associated delay time, we can characterize them as concerted or sequential trajectories. In Fig.~\ref{fig:sampletraj_delyatimes} we show  examples of  a concerted and several  sequential trajectories. 

\begin{figure}
 \centering
 \includegraphics[width=0.95\textwidth]{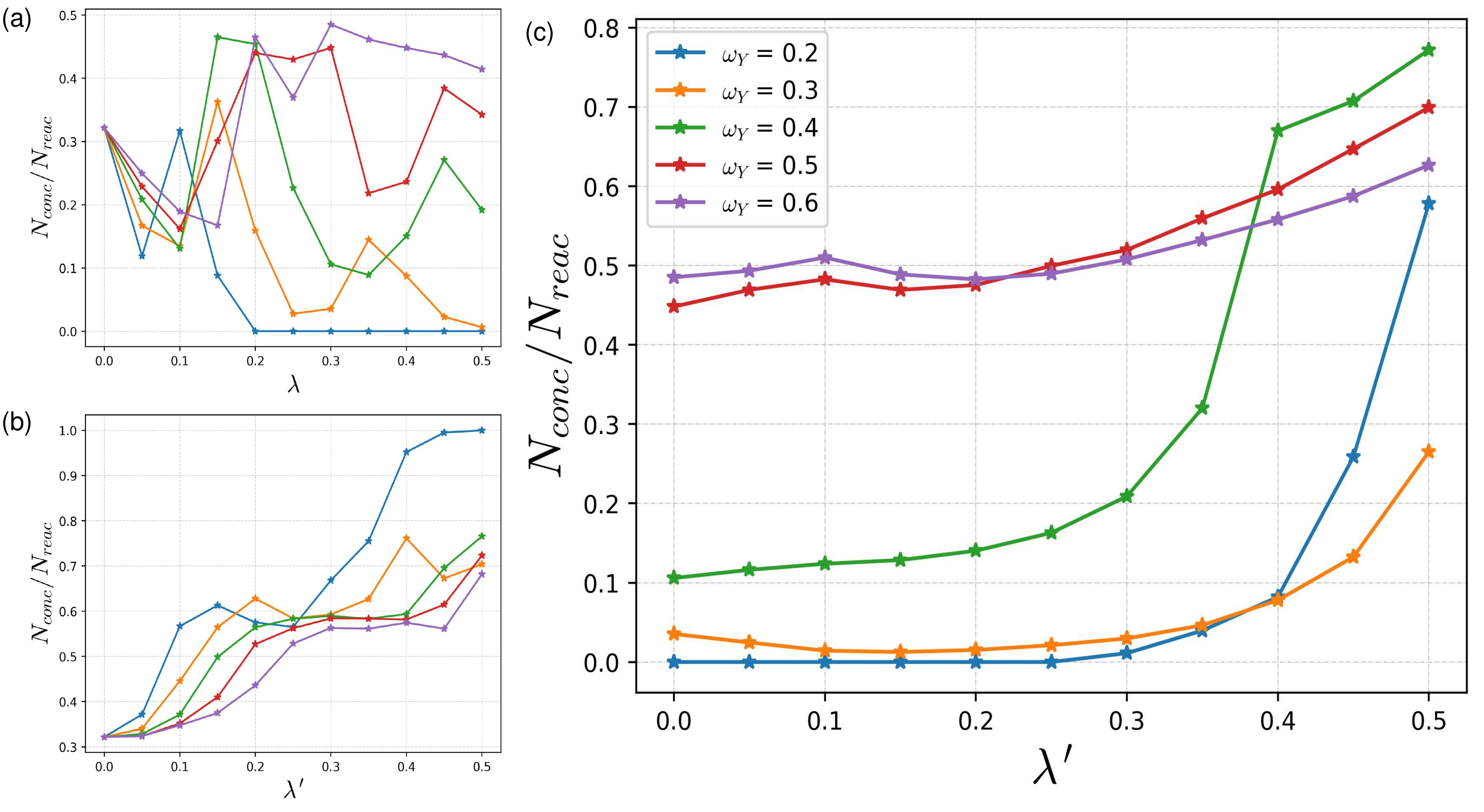}
 \caption{The fraction of concerted trajectories ($f_{conc} = N_{conc}/N_{reac}$) as a function of the coupling parameters. Here $N_{reac}$ is the total number of reactive trajectories i.e., ones that start at the index-$2$ saddle and go to the product P and reactant R regions in forward and backward time propagation. The total integration time is $t_{f} = 300$. (a) Variation with $\lambda$ for fixed $\lambda' = 0$ (b) Variation with $\lambda'$ for fixed $\lambda = 0$ (c) Variation with $\lambda'$ for fixed $\lambda = 0.3$. The frequency $\omega_{Y}$ of the transverse $Y$-mode are shown in the legend. It is important to note that $N_{reac}$ varies with the $(\lambda,\lambda',\omega_Y)$ parameters.}
 \label{fig:fraction_conc}
\end{figure}

For the delay time computation we adopt the following strategy. The first instance when the forward time trajectory enters either of the intermediate regions $I_{1,2}$ is noted as $t_I$. This event corresponds to the transfer of either one of the proton (cf. Fig.~\ref{Fig1}). Subsequently, the time at which the trajectory enters the defined product region $P$ is denoted as $t_P$. This time corresponds to the second proton transfer and hence $\Delta \tau = t_{P} - t_{I}$ is associated as the delay time for the specific trajectory. Two points are important to note at this stage. First, between $t_I$ and $t_P$ the trajectory may visit the reactant region $R$ or visit the intermediate regions several times. Second, the $\Delta \tau$ as defined is sensitive to the extent of the different regions. Thus, variations in the region sizes can change the $\Delta \tau$ for a given trajectory. Examples for the same are shown in Fig.~\ref{fig:sampletraj_delyatimes} and it is clear that some of the large $\Delta \tau$ can become considerably smaller or a short $\Delta \tau$ sequential trajectory can turn into a concerted trajectory. However, the effect of such variations on the distribution $P(\Delta \tau)$ shown in Fig.~\ref{fig:histogram} is not expected to be significant. In particular, the qualitative trends seen in Fig.~\ref{fig:fraction_conc} and Fig.~\ref{fig:histogram} are robust to small variations in the region sizes.

\subsection{Effect of the third degree of freedom on the concerted pathways}
\label{subsect:fconc}

Before discussing our results for the delay time distributions, in Fig.~\ref{fig:fraction_conc} we show the influence of the third degree of freedom coupling in Eqn.~\ref{massscaledham3D} on the ``pure" (as opposed to dynamically) concerted mechanism. In particular, Fig.~\ref{fig:fraction_conc}(a) and (b) show the fraction of concerted trajectories $f_{conc}$ upon coupling only the $X_s-Y$ modes ($\lambda' = 0)$ and  the $X_a-Y$ modes ($\lambda = 0$) respectively. Note that the Hamiltonian in Eqn.~\ref{massscaledham3D} involves both the couplings and hence the results in Fig.~\ref{fig:fraction_conc}(a) and (b) are a bit artificial. Nevertheless, such an analysis allows for dissecting, and a better understanding, of the results for the actual $a_g$-symmetry coupling form. It is clear from Fig.~\ref{fig:fraction_conc}(a) that, apart from the initial oscillatory nature\footnote{We remark here that the $f_{conc}$ in Fig.~\ref{fig:fraction_conc}(a) exhibits peaks at certain values of $\lambda \equiv \lambda_p$. Interestingly, these peaks seem to occur when the reactant well harmonic frequencies $\omega_{s}(\lambda)$ and $\omega_{a}$ become degenerate. An approximate estimate is 
$\lambda_p \approx \omega_Y \left[(\omega_{a}/2 \Delta X_{s})^{2} - 2 \bar{\alpha}_{s} \right]^{1/2}$. For the parameters of interest, $\lambda_p \approx 0.47 \omega_Y$. Note that the degeneracy is driven by the $X_{s}-Y$ coupling and hence an explcit three degrees of freedom effect. At the present moment we do not have a dynamical insight into this observation.}, low  $\omega_{Y}$ tend to drastically reduce $f_{conc}$ for increasing $\lambda$, while relatively larger $\omega_{Y}$ lead to a slight increase.  On the other hand, the results in Fig.~\ref{fig:fraction_conc}(b) indicate that $f_{conc}$ increases moderately upon increasing the $X_{a}-Y$ mode coupling strengths. 
Therefore,  the full $a_g$-symmetry coupling case with $\lambda,\lambda' \neq 0$ should encode the subtle competition between the two couplings. This is confirmed in Fig.~\ref{fig:fraction_conc}(c) where, as an example, the variation in $f_{conc}$ with $\lambda'$ for a fixed value of $\lambda = 0.3$ is shown. Interestingly, now the high $\omega_{Y}$ cases show very little variation over a significant range of the $X_{a}-Y$ coupling strengths. In contrast, for $\omega_{Y} \leq 0.4$  the results are more complex with $f_{conc}$ increasing  with  $\lambda'$ and the oscillations seen in Fig.~\ref{fig:fraction_conc}(a) being absent. As expected, for $\lambda' > \lambda = 0.3$ one observes $f_{conc}$ increasing substantially. Nevertheless, it is evident from Fig.~\ref{fig:fraction_conc}(c) that even for relatively large $\lambda'$ values the $\omega_Y = 0.2$ and $0.3$ cases have considerably lower $f_{conc}$  in comparison to the uncoupled case. We remark that these results agree with the general expectation that coupling of the large amplitude (low frequency) modes to the reaction coordinate can lead to dynamical behaviours that are vastly different from the dynamics of reduced dimensional systems. 

 Note that Fig.~\ref{fig:fraction_conc} pertains to the pure concerted pathways and hence, by definition, zero delay times. Based on the discussions in the introduction, a useful perspective is to focus on the fraction of dynamically concerted trajectories. Thus, although Fig.~\ref{fig:fraction_conc} indicates that low values of $\omega_Y$ lead to a reduced $f_{conc}$, is it possible that most of the trajectories are still dynamically concerted for a reasonable choice for the delay time cutoff $\Delta \tau_c$. In other words, if the distribution of delay times $P(\Delta \tau)$ associated with the initial ensemble of trajectories in Fig.~\ref{fig:fraction_conc} is strongly peaked for $\Delta \tau \leq \Delta \tau_c$ then the mechanism would be labeled as dynamically concerted. Consequently, as argued by Black et al. ~\cite{black2012dynamics}, the significant lowering of $f_{conc}$ for small $\omega_Y$ values observed in Fig.~\ref{fig:fraction_conc} need not really imply a major change in the reaction mechanism. Therefore, to ascertain if this indeed is the case we now turn our attention to the computation of the delay time distributions. 
 
 \subsection{Delay time distributions: importance of the low frequency transverse modes}
\label{subsect:delaytime}

From the discussions above, it is clear that in order to analyze $P(\Delta \tau)$ results for our model system it is essential to define the cutoff $\Delta \tau_c$. 
One possible choice for this cutoff time is related to the lifetime of a TS according to the Eyring equation~\cite{horn1996retro,xu2010dynamics}. This timescale is set by the prefactor of TST i.e., $\Delta \tau_{c} \sim h/k_{B} T$ with $h, k_{B}$, and $T$ being the Planck constant, Boltzmann constant and temperature respectively. However, as we are dealing with a index-$2$ saddle and the dynamics is at zero temperature, we choose $\Delta \tau_c$ based on the timescales associated with the reactant oscillations~\cite{black2012dynamics} or the unstable motion at the index-$2$ saddle. Such criteria have been invoked before in several studies~\cite{yang2019mechanisms}. For our model system and parameters of interest, as shown in \ref{jacobian_vf}, the unstable frequencies $\Omega_{s}^{*} \sim 1.6$ and $\Omega_{a}^{*} \sim 1.3$ at the index-$2$ saddle. These frequencies, independent of the couplings $(\lambda,\lambda')$ due to the form of the Hamiltonian, translate roughly to a timescale $T^{*}_{s,a} \sim 5$. On the other hand, of the two harmonic frequencies around the reactant minimum, only $\Omega_{s}$ depends on $(\lambda,\omega_Y)$ and varies from $\sim 5.5$ ($\omega_Y=0.2$) to $\sim 3.0$ ($\omega_Y = 0.5$), while $\Omega_{a} \sim 2.7$ stays fixed. Consequently, the harmonic timescales associated with the proton transfer modes at the reactant minimum is about $T_{h} \sim 2$. In this work, we therefore choose the conservative estimate $\Delta \tau_c \sim 5$ for discussing the delay time results.  

\begin{figure}
 \centering
 \includegraphics[width=1.0\textwidth]{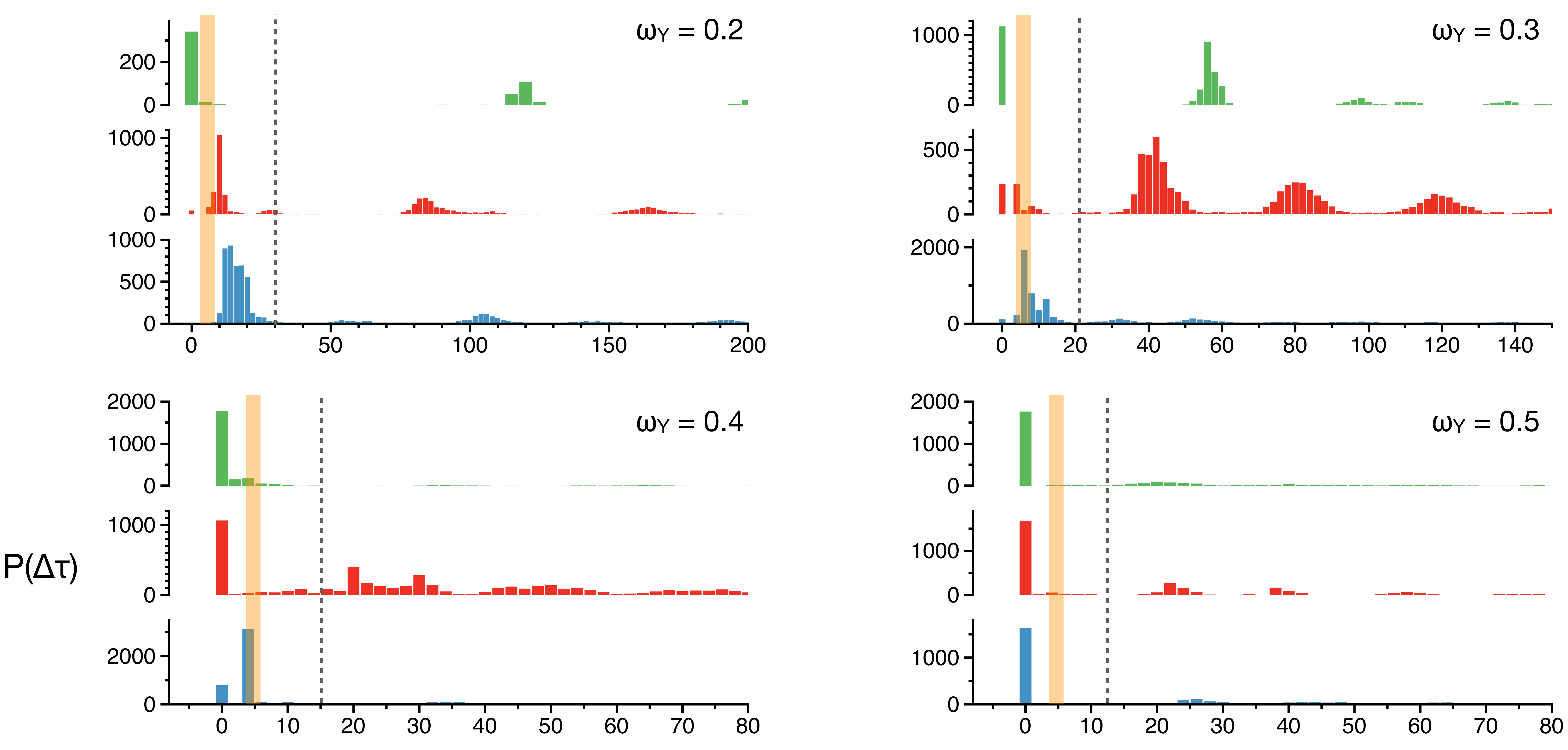}
 \caption{Delay time distributions $P(\Delta \tau)$ for fixed $\lambda = 0.3$ and varying $\lambda'$. The frequency $\omega_Y$ of the transverse Y-mode is indicated in each case. The histograms in blue, red, and green correspond to $\lambda'$ value $0.1$, $0.3$ and $0.5$ respectively. The orange vertical bar at $\Delta \tau_{c} \sim 5$ indicates the timescale associated with the unstable directions at the index-$2$ saddle. Note that the axis scales are different for each case and for comparison in each case the vertical dashed line corresponds to the harmonic period $2 \pi/\omega_Y$ associated with the transverse mode.}
 \label{fig:histogram}
\end{figure}

The results of the delay time computations are shown in Fig.~\ref{fig:histogram} for fixed $\lambda = 0.3$ as histograms\footnote{Note that as discussed previously and shown in Fig.~\ref{fig:sampletraj_delyatimes}, changing the region sizes will lead to some reshuffling of the counts, particularly for those with very large delay times. Nevertheless, the small to moderate time counts and their observed shifts should be robust.}. In each panel of Fig.~\ref{fig:histogram} the transverse mode frequency $\omega_Y$ is fixed and the delay time distributions for three values of $\lambda'$ are shown. Note that the both $f_{conc}$, shown in Fig.~\ref{fig:fraction_conc}, and $P(\Delta \tau)$ are computed using same initial ensemble. In addition, the parameters used for generating Fig.~\ref{fig:histogram}  are fairly representative of other parameter sets as well. For the value of $\omega_Y = 0.4$ and $0.5$ Fig.~\ref{fig:histogram} (bottom panels) shows that a large fraction of the distribution is concentrated for $\Delta \tau \leq \Delta \tau_c$, implying that  the mechanism is  dynamically concerted. In contrast, for the case of $\omega_Y = 0.2$  it is clear that the mechanism is sequential for $\lambda' = 0.1$ and $0.3$ with the emergence of dynamically concerted behaviour for larger coupling strengths. However, the fact that there are substantial peaks for $\Delta \tau \gg 10$  does hint at a fairly complex reaction dynamics. Clearly, the most complex variations in the distribution are seen in Fig.~\ref{fig:histogram} for the $\omega_Y = 0.3$ case. Here, despite the general trend of the onset of dynamical concerted behaviour with increasing $\lambda'$, even for the largest coupling a substantial fraction of the trajectories exhibit dynamically sequential mechanism. Given the opposing trends in $f_{conc}$ observed in Fig.~\ref{fig:fraction_conc}(a) and (b), one perhaps anticipates the $\lambda \sim \lambda'$ case for lower values of $\omega_Y$ to be in a sort of ``crossover" region. 

It is worthwhile pointing out the following interpretation of the delay time distribution results presented here. In a given molecular system, characterized by the parameters $G$ and $D$, the multitude of $a_g$-symmetry modes couple with a range of $\omega_Y, \lambda$, and $\lambda'$ values.  The results in Fig.~\ref{fig:histogram} then suggest that in the full multidimensional system whether the mechanism is dynamically concerted or sequential depends rather sensitively on the set of ratios $[(\lambda/\lambda')_1,(\lambda/\lambda')_2,\ldots,(\lambda/\lambda')_{n_{ag}}]$, where $n_{ag}$ is the total number of $a_g$-symmetry modes in a specific molecule. Clearly, similar criteria should exist for other transverse modes belonging to different symmetry classes. At the moment there is not much known about the dynamical competition between two or more low frequency modes with different symmetries. Nevertheless, Fig.~\ref{fig:histogram} does provide a clue as to why any a priori decision on the mechanism based solely on the static PES features is bound to be problematic. To this end, in the following section we provide further support by establishing a link between the phase space manifolds and the delay time distributions.

\subsection{Phase space viewpoint: Lagrangian descriptors are correlated with delay time distributions}
\label{subsect:LD_delaytime}

A crucial observation, as shown in~\ref{jacobian_vf}, is that the linear analysis of the index-2 saddle equilibrium does not shed any light on the changes in the fraction of concerted trajectories with coupling strengths shown in Fig.~\ref{fig:fraction_conc}. The eigenvalues of the linearized system at the index-2 saddle is independent of the coupling strengths, $\lambda, \lambda^\prime$. This implies that the competition of concerted vs sequential pathways for initial conditions launched from the vicinity of the index-2 saddle is inherently mediated by the global phase space structures. More so, these are global invariant manifolds in the phase space and transport initial conditions between intermediate and product wells. Thus, differentiating which initial conditions have low delay time, that is dynamically concerted, and high delay time, that is dynamically sequential. 

\begin{figure}
    \centering
    \includegraphics[width=1.0\textwidth]{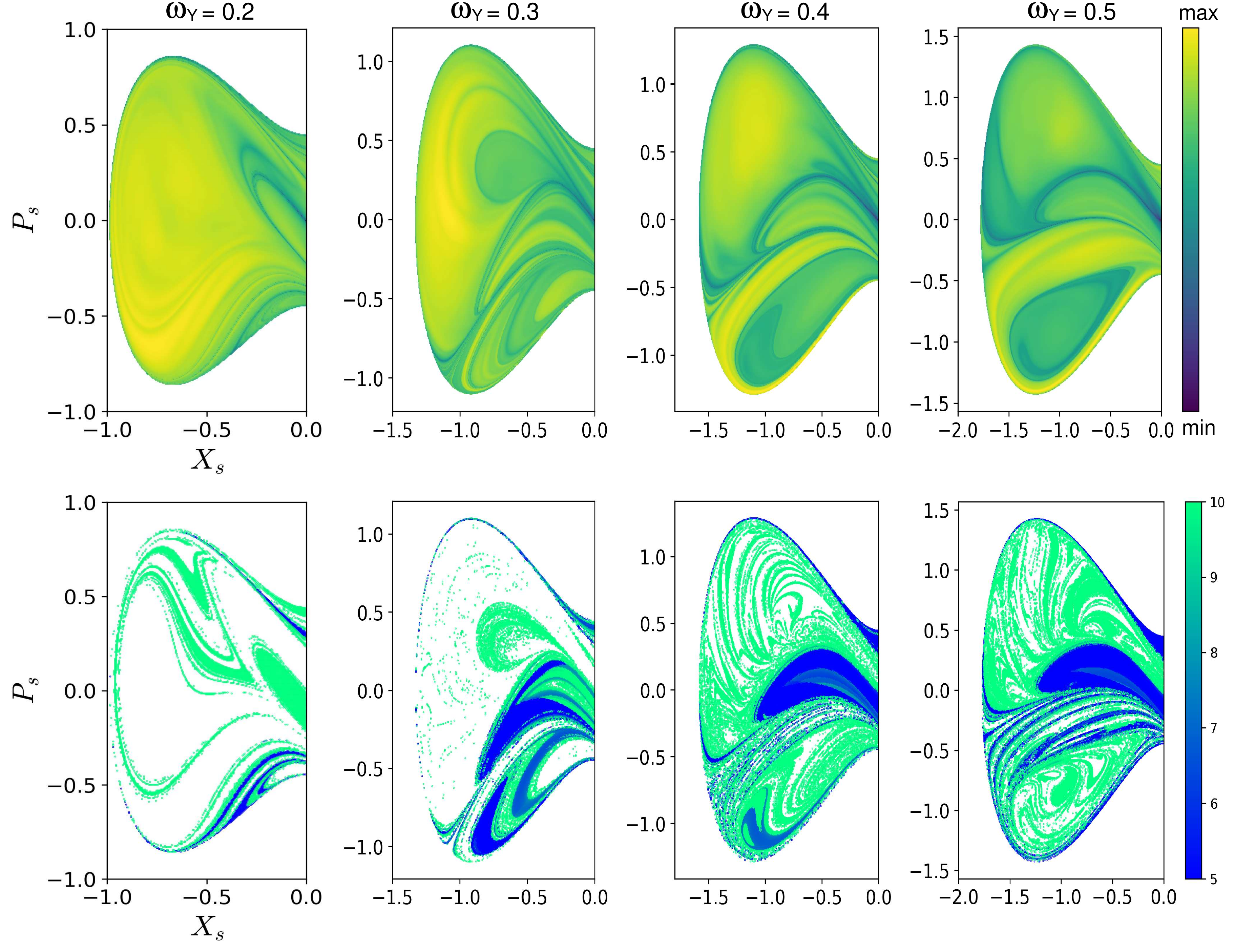}
    \caption{(Top panel) Forward LD map for $\tau = 10$ and (bottom panel) delay time map in the ($P_s,X_s)$ space at total energy of $E = 1.1$. The parameter values are  $\lambda = 0.3$, $\lambda^\prime = 0.1$. The frequency $\omega_Y$ of the transverse Y-mode is indicated above each column. The color scales associated with LD and delay time map are indicated in the respective panels. Note that the initial conditions of ``pure" concerted and dynamically concerted trajectories are shown in blue. The green color indicates the initial condition of trajectories with delay time $10$ or greater in the delay time map. The empty (white color) spaces in the delay time map correspond to initial conditions that are non-reactive up to the final time of integration.}
    \label{fig:Ld_delya_combined_0.1}
\end{figure}

In this study, we use Lagrangian descriptors~\cite{mancho_lagrangian_2013,lopesino_theoretical_2017,ld_book_2020} (see~\ref{sect:ld_theory} for details on the method) to identify the changes in the phase space structures with the changes in the coupling strengths and frequency of the third mode. In the case of three or more degrees of freedom systems, this method has been used to detect invariant manifolds and reactive islands~\cite{naik_finding_2019a,naik_finding_2019b,naik_detecting_2020}, discovering structure in the nuclear phase space in nonadiabatic quantum dynamics~\cite{eklund_investigating_2021}, while there is an increasing number of analysis for one and two degrees of freedom system with and without dissipation and time dependence. We refer the reader to the references in the open-source book on Lagrangian descriptors~\cite{ld_book_2020}. However, for three degrees of freedom systems with multiple saddles with varying indices, the use of LD has not been studied carefully and we present some preliminary discussion of this method.

\begin{figure}
    \centering
    \includegraphics[width=1.0\textwidth]{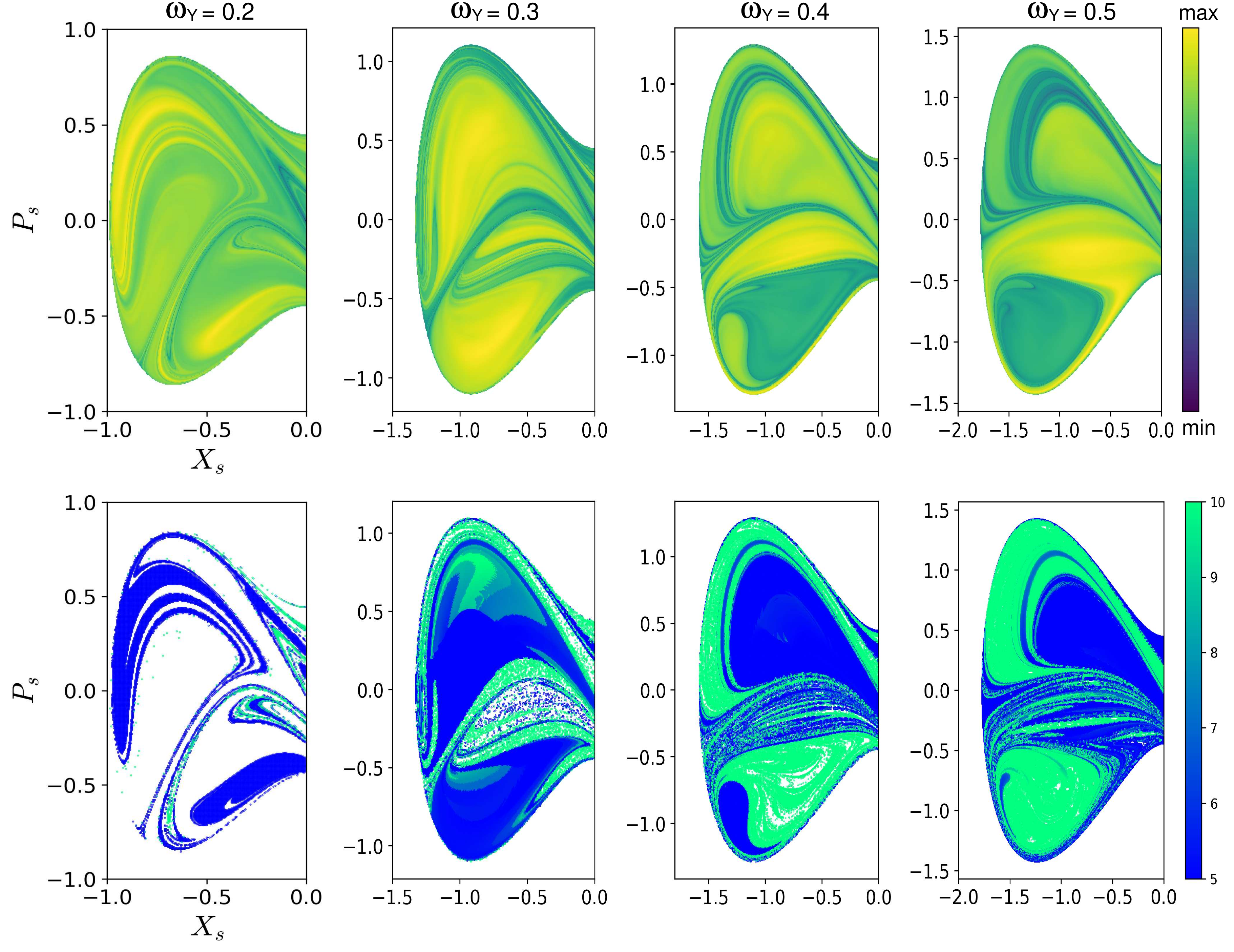}
    \caption{Same as in Fig.~\ref{fig:Ld_delya_combined_0.1} with parameter values  $\lambda = 0.3$ and $\lambda^\prime = 0.3$.}
    \label{fig:Ld_delya_combined_0.3}
\end{figure}

\begin{figure}
    \centering
    \includegraphics[width=1.0\textwidth]{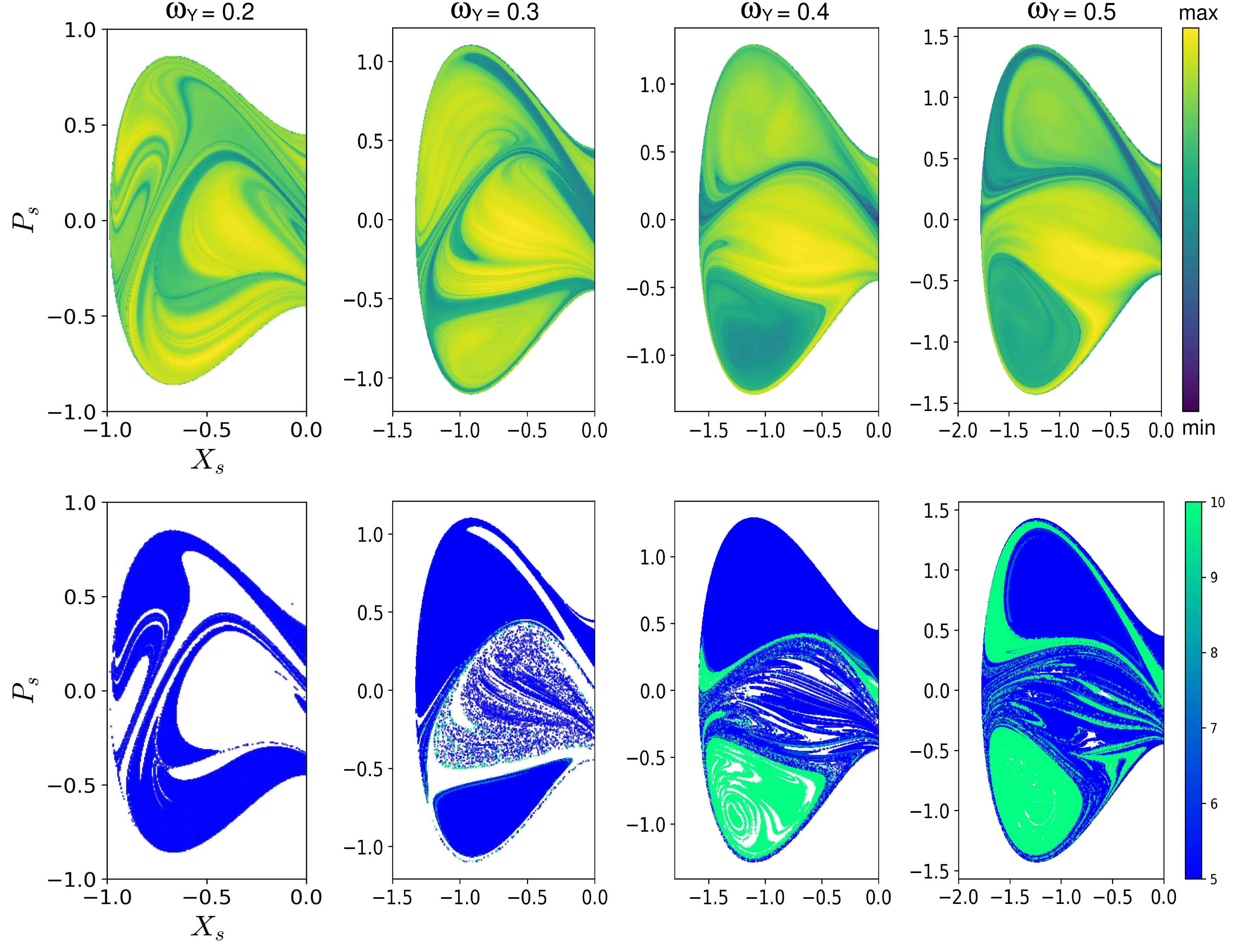}
    \caption{Same as in Fig.~\ref{fig:Ld_delya_combined_0.1} with  parameter values  $\lambda = 0.3$ and $\lambda^\prime = 0.5$.}
    \label{fig:Ld_delya_combined_0.5}
\end{figure}

For the three degrees of freedom system, we define the two dimensional section on the five dimensional energy surface
\begin{equation}
    \Sigma_{X_sP_s}^+ = \left\{ (X_s, X_a, Y, P_s, P_a, P_y) \in \mathbb{R}^6 \, | \, X_a = 0, Y = 0, P_y = 0, \dot{X}_a > 0 \right\}
    \label{eqn:2dsection_3dof}
\end{equation}
to inspect changes in the phase space structures with changes in the coupling strength and frequency of the third mode. We compute the Lagrangian descriptor for initial conditions on the reactant side ($X_s \leq 0$) of the section (Eqn.~\eqref{eqn:2dsection_3dof}). The integration time used if $10$ time units which is almost double the cut-off time for dynamical concerted ($\Delta \tau_c \sim 5$) pathway. While most studies using the LD method, and supported by theoretical arguments, tend to choose high integration time, we found that the short integration time of $10$ time units gave sufficient time for the structures to form and did not generate the many intricate stretching and folding of the global invariant manifolds. We compare the LD contour map with the delay time map in Figs.~\ref{fig:Ld_delya_combined_0.1}-\ref{fig:Ld_delya_combined_0.5} to show, for the first time, a striking correspondence between the invariant manifolds and delay time distribution. First, we observe that a direct correspondence in the contours of delay times and LD values across all the coupling strengths and frequency of the third mode with a cut-off time for dynamically concerted behaviour of $\Delta \tau_c \sim 5$ time units. The invariant manifolds identified in the LD contour map correspond to initial conditions with high delay time. However, the regions bounded by the invariant manifolds have two distinct delay times, that is either below $\Delta \tau < 5$ or $\Delta \tau > 10$. It implies that the invariant manifolds partition the initial conditions into dynamically concerted or sequential mechanisms. However, it is unclear as to which invariant manifolds can be unambiguously tied to a given mechanism. In order to discern which invariant manifolds mediate dynamically concerted and dynamically sequential mechanisms for energies above the index-2 saddle, one needs to evolve an ensemble of trajectories inside regions bounded by the invariant manifolds. This needs to be paired with a computation of the normally hyperbolic invariant manifolds~\cite{wiggins_normally_2014} and it's associated global invariant manifolds. The geometry of the normally hyperbolic invariant manifolds (3-sphere) associated with the index-2 saddle and its stable and unstable invariant manifolds (spherical cylinders or with geometry $\mathbb{S}^2 \times \mathbb{R}$) is still an area of continued interest~\cite{ezra_phasespace_2009,collins2011index,pandey2021classical} and we expect their structure and stability in the parameter space will shed light on the precise phase space mechanism of the competition between the concerted and sequential pathways.

\section{Conclusion and Outlook}
\label{sect:conc}

In this work we have studied the classical dynamics of a  three degrees of freedom Hamiltonian which models the double proton transfer reaction in a particular class of molecules. However, the analysis and techniques presented here are expected to be relevant for other types of systems which involve  breaking and forming of multiple bonds. The key points that emerge from our study are as follows:
\begin{enumerate}
    \item Coupling of additional low frequency modes to the reactive modes can lead to a change in the reaction mechanism inferred from lower dimensional studies. In particular, in the context of the DPT reaction studied in this work, it is seen that even a single low frequency mode can substantially change the fraction of reactive trajectories that proceed along the concerted pathway. It would be instructive to construct the phase space dividing surface for the index-$2$ saddle, along the lines of the earlier work by Collins, Ezra, and Wiggins~\cite{collins2011index}, to gain further insights into the  modulation of the fraction of concerted trajectories.
    \item Inspired by several earlier studies on various reactions that involve multiple bond formation, we have explored the utility of classifying reactions as dynamically concerted or sequential. To this end a simple, but dynamical, measure involving the time delay between the formation of two bonds was used. We suggest that the distributions of delay times provides much more information then the fraction of concerted trajectories and its use is motivated by the work of~\cite{black2012dynamics}. More importantly, we show a direct correspondence between the delay times and the phase space invariant manifolds for the coupling strengths and frequency of the third mode. This observation, therefore, places the earlier studies on a firm dynamical basis.
    \item We have shown that the technique of Lagrangian descriptors can be invoked to map out the relevant invariant manifolds in high dimensional phase space. In particular, although not  explored further in the current work, we observe that the LDs do encode the manifolds responsible for both the dynamically concerted and the dynamically sequential reactive pathways. Further studies on the changes in the structure of the index-$2$ normally hyperbolic invariant manifolds (NHIM)~\cite{wiggins2016role,ezra2009phase} with the transverse mode frequency and connection to the LD maps will be the focus of our future work.    
\end{enumerate}

Several issues arise in the context of our, admittedly preliminary, study and we briefly mention a few. Firstly, are the concerted and sequential pathways uncorrelated? One way to address this is to compute the so called gap time distribution~\cite{slater1956formulation,slater1959theory,ezra2009microcanonical,thiele1962comparison,thiele1963comparison} for the model system and the possible connections to the delay time distributions. Such a connection, along with an unambiguous disentangling of the phase space invariant manifolds for the two mechanisms, will then allow for decomposing the rate of the reaction in terms of ``concerted rates" and ``sequential rates". Secondly, the extent of  intramolecular vibrational energy redistribution (IVR)~\cite{leitner2015quantum,keshavamurthy2013scaling,nesbitt1996vibrational} amongst the modes  needs to be brought out clearly. Since the model has three degrees of freedom, it would be relevant to map out the Arnold web structure~\cite{karmakar2018relevance,karmakar2020intramolecular,karmakar2020stable} in the intermediate wells and correlate with the residence time distributions~\cite{shojiguchi2007fractional}. Such insights from the IVR dynamics may lead to the identification of the ``trigger" modes of the molecule that ultimately result in a concerted or sequential mechanism. Finally, our entire study is classical and raises the question of whether the quantum dynamics also allows for a dynamically concerted or sequential classification. The issue is subtle since, apart from tunneling which is relevant at low temperatures and correlated dynamics due to quantum entanglement~\cite{smedarchina2018entanglement}, even a proper definition of the delay time may pose difficulties.

We conclude by noting that the Hamiltonian in eq.~\ref{massscaledham3D} leads to a very rich and complex dynamics. Our study here has explored only a thin ``slice" of the vast parameter range.  We hope that a more detailed phase space analysis and classical-quantum correspondence study of the model presented herein will lead to further insights into the dynamical implications of high index saddles on reaction mechanisms. 

\section{Acknowledgement}
We thank Antonio Fern\'{a}ndez-Ramos and Zorka Smedarchina for clarifying certain aspects of the potential energy surface used in their various studies of DPT. SK's research is supported by the Science and Engineering Research Board (SERB) India (project no. EMR/006246). SN acknowledges the support of EPSRC Grant No. EP/P021123/1 (CHAMPS project). PP thanks the IIT Kanpur for graduate fellowship and the High Performance Computing Facility at IIT Kanpur for computing resources.


\appendix
\section{Double proton transfer Hamiltonian: scaled $2$D model}
\label{1DHamiltonian}

For a single proton transfer process, labeled as subsystem $1$ in Fig.~\ref{Fig1}, the standard model corresponds to a quartic double well oscillator with the Hamiltonian
\begin{equation}
    \bar{H}(\bar{x},\bar{p}) = \frac{1}{2\bar{m}} \bar{p}^{2} + U(\bar{x})
\end{equation}
and the potential energy function
\begin{equation}
    U(\bar{x}) = -a \bar{x}^{2} + b \bar{x}^{4} + U_{0}
\end{equation}
with $\bar{a},\bar{b} > 0$ and $U_{0}$ being the barrier height for the single barrier proton transfer. The critical points of the potential are determined as $\bar{x}_{c} = 0, \pm \Delta x_{0}$ with $\Delta x_{0} \equiv (a/2b)^{1/2}$. The point $x_{c} = 0$ corresponds to the maximum with $U(0) \equiv U_{0} = a^{2}/4b$ while $x_{c} = \pm \Delta x_{0}$ are the two minima with $U(x_c = \pm \Delta x_{0}) = 0$. Thus, $U_{0}$ is the barrier height. As indicated in Fig.~\ref{Fig1}, the distance between the two minima is equal to $2 \Delta x_{0}$.

We introduce scaled variables as follows. The coordinate is scaled  by $\Delta x_{0}$ as $\bar{x} = x \Delta x_{0}$, with  $x$ being dimensionless. Thus, the potential energy transforms as 
\begin{eqnarray}
U(x) &=& -a (\Delta x_{0})^{2} x^{2} + b (\Delta x_{0})^{4} x^{4} + U_0 \\
     &=& U_0 \left[x^{2}-1 \right]^{2}
\end{eqnarray}
Consequently, the Hamiltonian can be written down as
\begin{equation}
    \bar{H}(x,P) = \frac{(\Delta x_0)^{2}}{2\bar{m}} P^{2} + U_0 \left[x^{2}-1 \right]^{2}
\end{equation}
where $P \equiv \bar{m} \dot{x}$. Measuring mass in units of the proton mass $m_{H}$ we have $\bar{m} = m_{H} m$ and $P = \bar{m} \dot{x} = m_{H} (m \dot{x}) \equiv m_{H} \tilde{p}$. Finally, scaling the energy by $2U_0$ and  time as $t = \alpha \tau$ with $\alpha = [(\Delta x_0)^{2} m_H/2U_0]^{1/2}$ we obtain the transformed dimensionless Hamiltonian
\begin{equation}
    H(x,p)  =  \frac{1}{2m} p^{2} + \frac{1}{2} \left[x^{2}-1 \right]^{2}
    \label{eq:1DscaledHam}
\end{equation}
with the identification $p \equiv m (dx/d\tau)$. 

We now consider the double proton transfer scenario shown in Fig.~\ref{Fig1} wherein the system has two such equivalent protons tunneling sites. A two degree of freedom Hamiltonian can then be expressed in terms of the two proton coordinates ${\bf x} = (x_1,x_2)$ and their corresponding conjugate momenta ${\bf p} = (p_1,p_2)$
\begin{equation}
     H(\mathbf{x},\mathbf{p})=H_0(\mathbf{x},\mathbf{p})+U_{\rm coup}(\mathbf{x})  
\label{eq:hamiltonian}
\end{equation}
where, the zeroth-order Hamiltonian is generalized from Eqn.~\ref{eq:1DscaledHam} and of the form
\begin{equation}
    H_0(\mathbf{x},\mathbf{p}) =  \sum_{j=1,2} \left[\frac{1}{2m_{\rm j}} p_{j}^{2} + U_{0}({x_j}) \right]    
\end{equation}
with  
\begin{equation}
    U_{0}(x_j) = \frac{1}{2} \left[{x_{j}}^2 - 1 \right]^2
\end{equation}
The zeroth-order form is appropriate in the limit that the two protons being transferred are not correlated. However, typically, the two proton motions are coupled and general symmetry-based arguments indicate that the correct form of the coupling potential is given by
\begin{equation}
    U_{\rm coup}(\mathbf{x}) = -2G{x_1}{x_2} -Dx_{1}^{2}x_{2}^{2} - C(x_{1}^{3}x_{2} + x_{1}x_{2}^{3}) + \ldots
\end{equation}
Thus, in principle there are  couplings of all order between the two modes. However, as has been noted earlier, from a perturbative perspective the first two leading order terms in the above expansion for $U_{\rm coup}({\bf x})$ are sufficient to capture most of the essential dynamical features of the system. Therefore, in what follows 
we take $G, D \neq 0$ and ignore the higher order terms.

At this stage it is useful to switch  from the local coordinates used above to the normal mode coordinates $(x_s,x_a)$ with the transformation $(x_1,x_2)= (x_s+x_a,x_s-x_a)$. The Hamiltonian in this new representation is given by
\begin{equation}
    H(x_s,x_a,p_s,p_a) = \frac{1}{2M} (p_{s}^{2} + p_{a}^{2}) + \left(\frac{\delta m}{M^{2}}\right) p_s p_a + U(x_s,x_a)
    \label{eqn:unscaled_ham2dof}
\end{equation}
with $M \equiv m_1 + m_2$ and $\delta m = m_1 - m_2$. Note the presence of the momentum coupling term. This term vanishes when we are looking at the symmetric $m_1 = m_2$ cases, as in the present work. However, when considering singly substituted isotope case like $m_1 = m_{H}$ and $m_2 = m_{D}$, for instance, then $\delta m \neq 0$. The potential thus transforms into
\begin{equation}
    U(x_s,x_a) = \alpha_s \left[x_{s}^{2} - (\Delta x_s)^{2}\right]^{2} + \alpha_a \left[x_{a}^{2} - (\Delta x_a)^{2}\right]^{2} + 2 R x_{s}^{2} x_{a}^{2} + {\cal U}(G,D)
\end{equation}
where we have denoted $R = 3 + D$ and $\alpha_{s} = \alpha_{a} \equiv \alpha = 1-D$ with
\begin{subeqnarray}
 \Delta x_{s,a} &=& \sqrt{\frac{1\pm G}{1-D}} \\
 {\cal U}(G,D) &=& 1 - \alpha_s (\Delta x_s)^{4} - \alpha_a (\Delta x_a)^{4}
\end{subeqnarray}
As a final transformation, and preparation for the three degree of freedom Hamiltonian of interest to the current work, we transform to mass-weighted coordinates via  $X_{s,a} \rightarrow \sqrt{M} x_{s,a}$. We thus obtain the Hamiltonian 
\begin{equation}
    H({\bf X},{\bf P}) = \frac{1}{2} (P_{s}^{2} + P_{a}^{2}) + \left(\frac{\delta m}{M}\right) P_s P_a + U(X_s,X_a)
    \label{app:massscaledham2D}
\end{equation}
with the potential energy term
\begin{equation}
    U({\bf X}) = \bar{\alpha}_{s} \left[X_{s}^{2} - (\Delta X_s)^{2}\right]^{2} + \bar{\alpha}_{a} \left[X_{a}^{2} - (\Delta X_a)^{2}\right]^{2} + 2\bar{R} X_{s}^{2} X_{a}^{2} + {\cal U}(G,D)
    \label{app:massscaledpot2D}
\end{equation}
In the above we have denoted $({\bf X},{\bf P}) \equiv (X_s,X_a,P_s,P_a)$ with the parameters $\Delta X_{s,a} = \sqrt{M} \Delta x_{s,a}$, $\bar{\alpha}_{s} = \bar{\alpha}_{a} \equiv \alpha/M^{2}$,and $\bar{R} \equiv R/M^{2}$. The above Hamiltonian with $\delta m = 0$ corresponds to Eqn.~\ref{massscaledham2D} in the main article.


\section{Hamiltonian vector field and linear stability of equilibria}
\label{jacobian_vf} 

The Hamiltonian vector field is given by

\begin{equation}
	\begin{aligned}
		\dot{X}_s = &~P_s + \dfrac{\delta m}{M} P_a \\
		\dot{X}_a = &~P_a + \dfrac{\delta m}{M} P_s \\ 
		\dot{Y}	  = &~P_y \\
		\dot{P}_s = &~- 4 \left(\bar{\alpha_s} (X_s^3 - X_s(\Delta X_s)^2) + \bar{R} X_s X_a^2 \right) + 2 \lambda X_s \left(Y - \dfrac{\lambda}{\omega_Y^2}X_s^2 - \dfrac{\lambda^\prime}{\omega_Y^2}X_a^2 \right) \\
		\dot{P}_a = &~- 4 \left(\bar{\alpha_a} (X_a^3 - X_a(\Delta X_a)^2) + \bar{R} X_a X_s^2 \right) + 2 \lambda^\prime X_a \left(Y - \dfrac{\lambda}{\omega_Y^2}X_s^2 - \dfrac{\lambda^\prime}{\omega_Y^2}X_a^2 \right) \\
		\dot{P}_y = &~- \omega_Y^2 \left( Y - \dfrac{\lambda}{\omega_Y^2}X_s^2 - \dfrac{\lambda^\prime}{\omega_Y^2}X_a^2 \right)
	\end{aligned}
	\label{eqn:massscaled_vf3dof}
\end{equation}

This vector field has same number of equilibria as the two degrees of freedom model except that each equilibria also has a third coordinate. The equilibria, total energies, and their linear stability is summarised in the Table.~\ref{tab:eqpt_energy_stab_scaledham3dof},
where the locations of the index-$1$ saddles and their energy are given by
\begin{align}
\left(X_s^{\ddagger}, X_a^{\ddagger}, Y^{\ddagger} \right) & = \left( \pm \sqrt{\dfrac{\bar{\alpha_s} \bar{\alpha_a} (\Delta X_s)^2 - \bar{\alpha_a} \bar{R} (\Delta X_a)^2}{(\bar{\alpha_s} \bar{\alpha_a} - \bar{R}^2)}}, \pm \sqrt{\dfrac{\bar{\alpha_s} \bar{\alpha_a} (\Delta X_a)^2 - \bar{\alpha_s} \bar{R} (\Delta X_s)^2}{(\bar{\alpha_s} \bar{\alpha_a} - \bar{R}^2)}}, \dfrac{\lambda}{\omega_Y^2}(X_s^{\ddagger})^2 + \dfrac{\lambda^\prime}{\omega_Y^2}(X_a^{\ddagger})^2 \right) \\
H^{\ddagger} = & \dfrac{1}{M^2(\bar{R}^2 - \bar{\alpha_a} \bar{\alpha_s})} \left[  \bar{R}^2 \bar{\alpha_s}(\Delta X_s)^4 + \bar{R}^2 \bar{\alpha_a}(\Delta X_a)^4 ) - 2\bar{R} (\Delta X_s \Delta X_a)^2 \bar{\alpha_s} \bar{\alpha_a} \right] + \mathcal{U}(G,D).
\end{align}

The linear stability of the equilibria in Table~\ref{tab:eqpt_energy_stab_scaledham3dof} is determined by the eigenvalue problem associated with the linearized vector field, $\mathbb{J} \mathbf{v} = \beta \mathbf{v}$, where the Jacobian $\mathbb{J}$ is given by Eqn.~\eqref{eqn:jacobian_3dof}, $\beta, \mathbf{v}$ are the eigenvalues and eigenvectors of the Jacobian at the equilibrium point. For the reactant, product, and intermediate wells, the eigenvalues are of the form $\pm \omega_{c1}^e, \pm \omega_{c2}^e, \pm \omega_{c3}^e$. For the saddles at energy $H^{\ddagger}$, the eigenvalues are of the form $\pm \lambda^e, \pm \omega_{s1}^e, \pm \omega_{s2}^e$ which makes these index-1 saddles. For the saddle at energy $E_s = 1$, the eigenvalues are of the form $\pm \lambda_{s1}^e, \pm \lambda_{s2}^e, \pm \omega_{s3}^e$ which makes this an index-2 saddle. 

The Jacobian of the vector field, $\mathbb{J}(X_s, X_a, Y, P_s, P_a, P_y)$, is given by 

\begin{align}
	\begin{pmatrix}
		0 & 0 & 0 & 1 & \dfrac{\delta m}{M}  & 0\\
		0 & 0 & 0 & \dfrac{\delta m}{M} & 1 & 0\\
		0 & 0 & 0 & 0 & 0 & 1 \\
		-\dfrac{\partial^2 V}{\partial X_s^2} & -\dfrac{\partial^2 V}{\partial X_a \partial X_s} & -\dfrac{\partial^2 V}{\partial Y \partial X_s} & 0 & 0 & 0 \\
		-\dfrac{\partial^2 V}{\partial X_s \partial X_a} & -\dfrac{\partial^2 V}{\partial X_a^2} & -\dfrac{\partial^2 V}{\partial Y \partial X_a} & 0 & 0 & 0 \\
		-\dfrac{\partial^2 V}{\partial X_s \partial Y} & -\dfrac{\partial^2 V}{\partial X_a \partial Y} & -\dfrac{\partial^2 V}{\partial Y^2} & 0 & 0 & 0	
	\end{pmatrix},
    \label{eqn:jacobian_3dof}
\end{align}

where

\begin{align}
	-\dfrac{\partial^2 V}{\partial X_s^2} = & -4\left[ \bar{\alpha}_s (3X_s^2 - (\Delta X_s)^2) + \bar{R} X_a^2 \right] + 2 \lambda \left( Y - \dfrac{\lambda}{\omega_Y^2}3 X_s^2 - \dfrac{\lambda^\prime}{\omega_Y^2}X_a^2 \right) \\
		-\dfrac{\partial^2 V}{\partial X_a^2} = & -4\left[ \bar{\alpha}_a (3X_a^2 - (\Delta X_a)^2) + \bar{R} X_s^2 \right] + 2 \lambda^\prime \left( Y - \dfrac{\lambda}{\omega_Y^2} X_s^2 - \dfrac{\lambda^\prime}{\omega_Y^2}3 X_a^2 \right) \\
	-\dfrac{\partial^2 V}{\partial X_a \partial X_s} = & -\dfrac{\partial^2 V}{\partial X_s \partial X_a} =  -4 \left(2\bar{R}  + \dfrac{\lambda \lambda^\prime}{\omega_Y^2}\right)X_s X_a \\
	-\dfrac{\partial^2 V}{\partial Y \partial X_s} = & -\dfrac{\partial^2 V}{\partial X_s \partial Y} = \phantom{-}  2 \lambda X_s \\
	-\dfrac{\partial^2 V}{\partial Y \partial X_a} = & -\dfrac{\partial^2 V}{\partial X_a \partial Y} = \phantom{-} 2 \lambda^\prime X_a \\
	-\dfrac{\partial^2 V}{\partial Y^2} = & \phantom{-} -\omega_Y^2 
\end{align}

\begin{figure}[!ht]
    \centering
    \includegraphics[width = 01.0\textwidth]{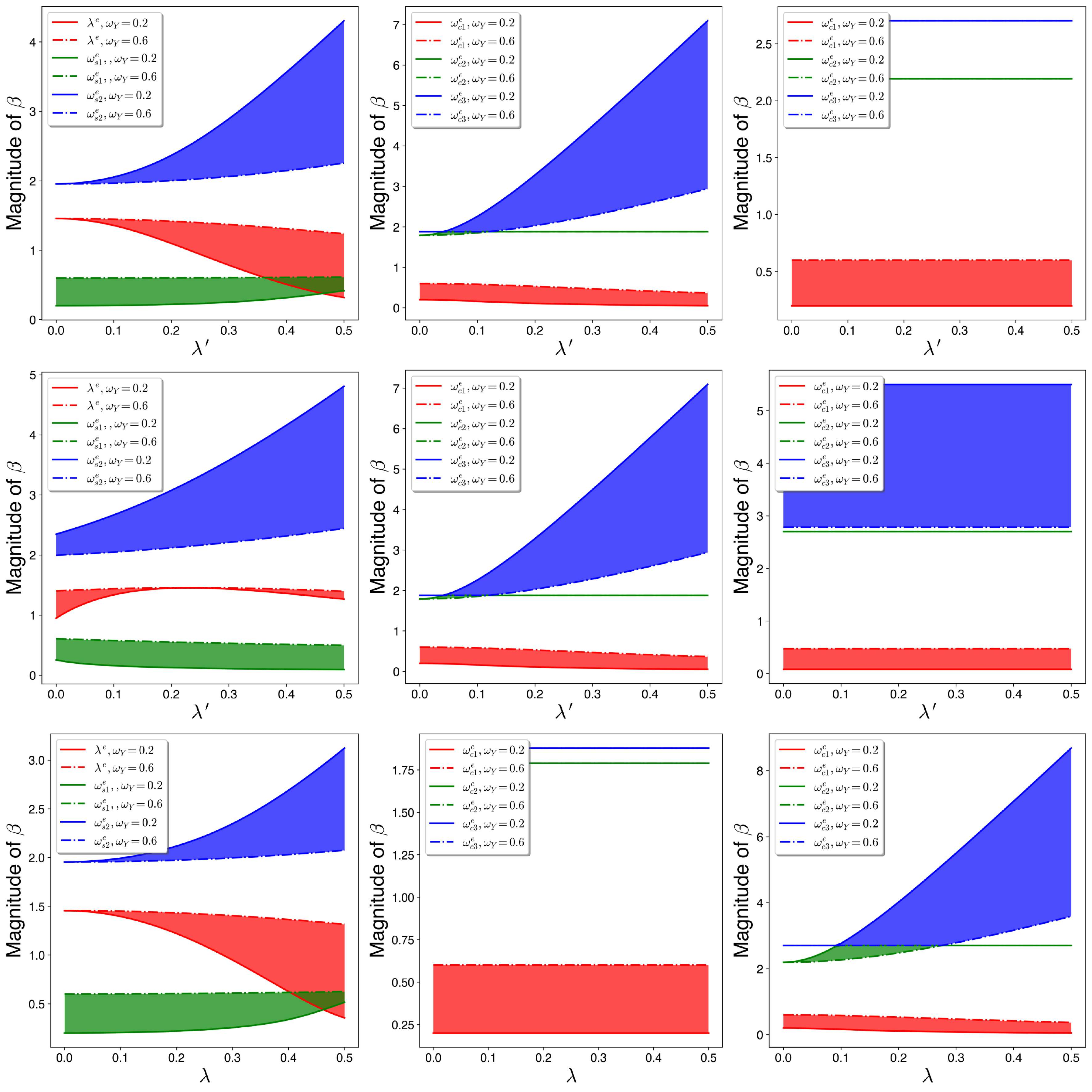}
    
    \caption{Eigenvalues variation with $\lambda^\prime, \lambda, \omega_Y$. (Top row) $\lambda = 0.0$ (Middle row) $\lambda = 0.3$ (Bottom row) $\lambda^\prime = 0.0$. Left, middle, and right columns correspond to the index-1 saddles, intermediate wells, and product (or reactant) well, respectively. The magnitude of the eigenvalues at $\omega_Y = 0.2$ and $\omega_Y = 0.6$ are shown as continuous line and dash-dot line, respectively, with the shaded region between the lines denoting the variation for $0.2 \leq \omega_Y \leq 0.6$. Other parameters are $\delta M = 0, M = 2, G = 0.20, D = 0.15$.}
    \label{fig:mag_eigenvalues}
\end{figure}

We track the changes in the linear stability of the index-1 and index-2 saddles with changes in the coupling strength by tracking the eigenvalues of the linearized vector field (Jacobian~\eqref{eqn:jacobian_3dof}) evaluated at the equilibrium points as the coupling strength is continuously varied. Note that in this work we have $\delta m = 0$.

The eigenvalues at the phase space point $(0,0,0,0,0,0)$  are given by
\begin{equation}
    \left[ \pm 2\sqrt{\bar{\alpha}_s} \Delta X_s, \pm 2\sqrt{\bar{\alpha}_a}\Delta X_a, \pm i \omega_Y  \right]
\end{equation}

which has the structure of an index-2 saddle and is only dependent on the parameters $D,G,C,\omega_Y$. This supports the parametric study in this work where the index-2 saddle maintains its linear stability as we vary $\lambda,\lambda^\prime$. Further, this also points to the fact that mere linear (local in the neighborhood of the equilibrium point) analysis will not reflect the dynamical mechanism due to the influence of the coupling parameters on the fraction of concerted trajectories. For the remaining equilibria, we show the changes in the magnitude of the eigenvalues along coupling parameters $\lambda^\prime$, $\lambda$ and third mode frequency $\omega_Y$ for $\lambda = 0.3$, $\lambda =  0.0$ and $\lambda^\prime = 0.0$ in Fig.~\ref{fig:mag_eigenvalues}. We observe that the eigenvalues of the intermediate wells are independent of $\lambda$ while the eigenvalues of the product (or reactant) well are independent of $\lambda^\prime$. In general, there are no critical changes in the eigenvalues as the coupling strengths and frequency of the third mode are varied. The stability type of the equilibria stays the same.

\section{Lagrangian descriptor: method to reveal the invariant manifolds}
\label{sect:ld_theory}

We briefly describe the method of Lagrangian descriptors, which reveals regions with qualitatively distinct dynamical behavior by showing the intersection of the invariant manifolds with the two dimensional section. For a general time-dependent dynamical system given by 
\begin{equation}
	\dfrac{d\mathbf{x}}{dt} = \mathbf{f}(\mathbf{x},t) \;,\quad \mathbf{x} \in \mathbb{R}^{n} \;,\; t \in \mathbb{R} \;,
	\label{eq:gtp_dynSys}
\end{equation}
where the vector field $\mathbf{f}(\mathbf{x},t)$ is assumed to be sufficiently smooth both in space and time. The vector field $\mathbf{f}$ can be prescribed by an analytical model or given from numerical simulations as a discrete spatio-temporal data set. For instance, the vector field could represent the velocity field of oceanic or atmospheric currents obtained from satellite measurements or from the numerical solution of geophysical models. For any initial condition $\mathbf{x}(t_0) = \mathbf{x}_0$, the system of first order nonlinear differential equations (given in Eqn.~\eqref{eq:gtp_dynSys}) has a unique solution represented by the trajectory that starts from that initial point $\mathbf{x}_0$ at time $t_0$.

\begin{figure}
    \centering
    \includegraphics[width = \textwidth]{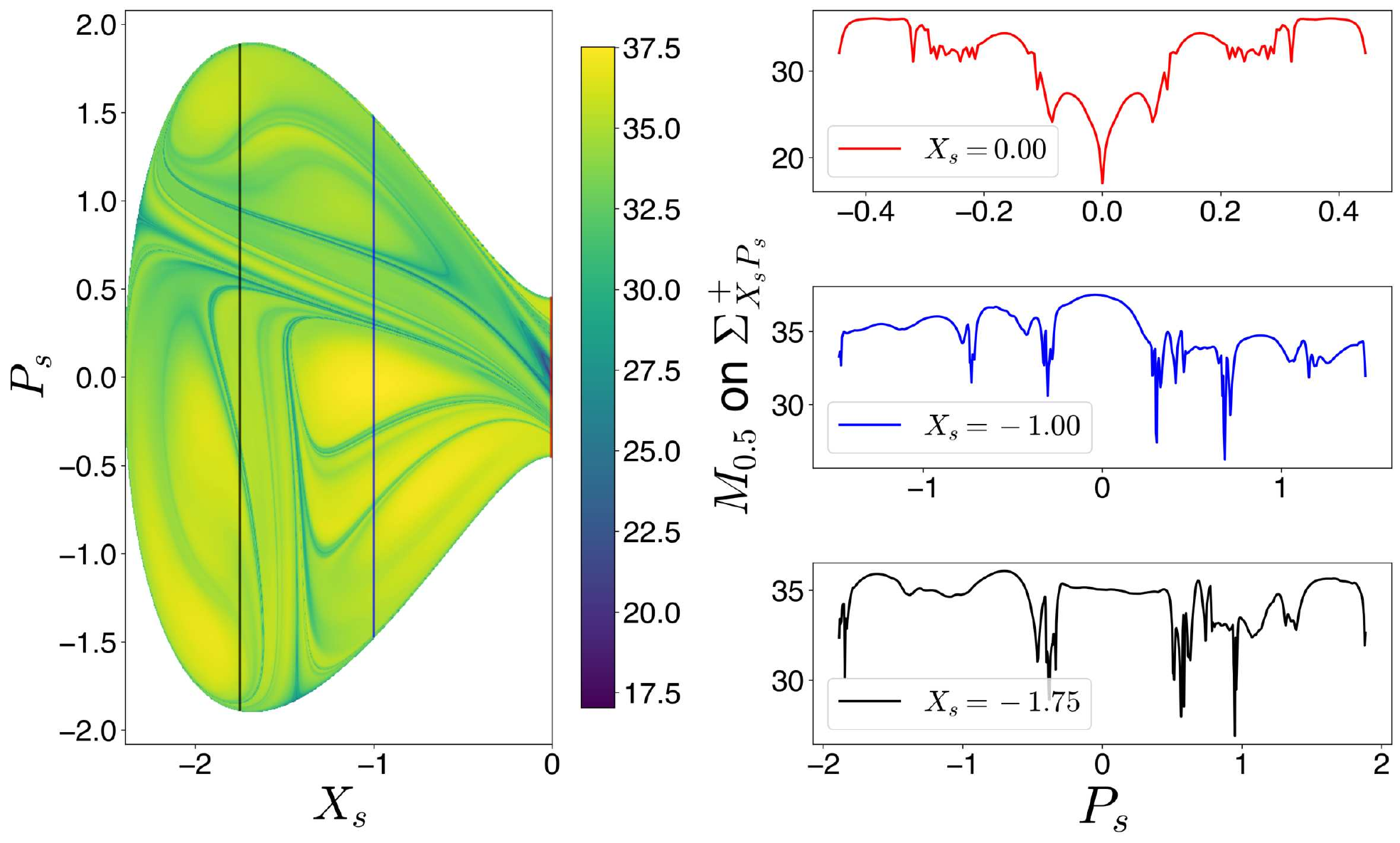}
    \caption{Lagrangian descriptor (forward) on the section~\ref{eqn:2dsection_3dof} identifying the invariant manifolds by the singular points with minima of the contour map shown by the one dimensional slices on the right. Other parameters are $\lambda = \lambda^\prime = 0, \omega_Y = 0.2$ and $E = 1.1$.}
    \label{fig:forwLD_uncoupled3dof}
\end{figure}

In this study, we adopt the LD definition
\begin{equation}
	\mathcal{L}_p(\mathbf{x}_{0},t_0,\tau) = \int^{t_0+\tau}_{t_0-\tau} \, \sum_{k=1}^{n}   \vert f_{k}(\mathbf{x}(t;\mathbf{x}_0),t) \vert^p \; dt  \;, \quad p \in (0,1]
	\label{eq:Mp_function}
\end{equation}
where $f_k$ is the $k-$the component of the vector field, Eqn.~\eqref{eq:gtp_dynSys} and use $p = 1/2$. We note that the integral can be split into its forward and backward time parts to detect the intersection of stable and unstable manifolds separately. This relates to finding the escape and entry channels into the potential well. In this study, we keep the forward part of the integral given by
\begin{equation}
	\begin{split}
		\mathcal{L}_p^{f}(\mathbf{x}_{0},t_0,\tau) & = \int^{t_0+\tau}_{t_0} \sum_{k=1}^{n} |f_{k}(\mathbf{x}(t;\mathbf{x}_0),t)|^p \; dt
	\end{split}
\end{equation}
Although this definition of LD does not have an intuitive physical interpretation as that of the arclength definition~\cite{mancho_lagrangian_2013}, it allows for a rigorous proof that the ``singular features" (non-differentiable points) in the LD contour map identify intersections with stable and unstable invariant manifolds~\cite{lopesino_theoretical_2017}. Another important aspect of what is known in LD literature as the
$p$-(quasi)norm is that degrees of freedom with relevance in escape/transition (reaction) dynamics can be decomposed and computed. This definition was used to show that the method can be used to successfully detect NHIMs and their stable and unstable manifolds in H{\'e}non-Heiles Hamiltonian~\cite{demian2017,naik2019b}. For this system, where both fixed (or variable) integration time is used, it has also been shown that the LD scalar field attains a minimum (or maximum) value along with singularity at the intersections of the stable and unstable manifolds, and given by
\begin{equation}
	\mathcal{W}^s(\mathbf{x}_{0},t_0) = \textrm{argmin } \mathcal{L}_p^{f}(\mathbf{x}_{0},t_0,\tau) \;,
	\label{eq:min_LD_manifolds}
\end{equation}
where $\mathcal{W}^s(\mathbf{x}_{0},t_0)$ are the stable manifolds calculated at time $t_0$ and  $\textrm{argmin}$ denotes the phase space coordinates on the two dimensional section that minimize the scalar field, $\mathcal{L}_p^{f}(\mathbf{x}_{0},t_0,\tau)$, over the integration time, $\tau$. Thus, the scalar field plotted as a contour map identifies the intersection of the stable manifold with a two dimensional section. This ability of LD contour map to partition trajectories with different phase space geometry is shown in the right panel of Fig.~\ref{fig:forwLD_uncoupled3dof} as singular values of LD identify the intersection of the manifolds with the chosen section. 

\bibliographystyle{elsarticle-num-names}
\bibliography{manuscript.bib}







\end{document}